\documentclass[aps,pra,twocolumn,superscriptaddress]{revtex4-1}

\usepackage{graphicx,enumerate,verbatim,bbold}
\usepackage{amsmath,amssymb,amsthm,float,mathrsfs}
\usepackage{dsfont}
\usepackage{braket}
\usepackage[colorlinks]{hyperref}
\usepackage[T1]{fontenc}
\usepackage[utf8]{inputenc}
\usepackage{lmodern}
\usepackage[caption=false]{subfig}

\begin{document}

\title{Two photons on an atomic beam-splitter: nonlinear scattering and induced correlations}


\author{Alexandre Roulet}
\affiliation{Centre for Quantum Technologies, National University of Singapore, 3 Science Drive 2, Singapore 117543, Singapore}

\author{Huy Nguyen Le}
\affiliation{Centre for Quantum Technologies, National University of Singapore, 3 Science Drive 2, Singapore 117543, Singapore}

\author{Valerio Scarani}
\affiliation{Centre for Quantum Technologies, National University of Singapore, 3 Science Drive 2, Singapore 117543, Singapore}
\affiliation{Department of Physics, National University of Singapore, 2 Science Drive 3, Singapore 117542, Singapore}

\date{\today}

\begin{abstract}
Optical emitters strongly coupled to photons propagating in one-dimensional waveguides are a promising platform for optical quantum information processing. Here, we present a theoretical study of the scattering of two indistinguishable photons on a single two-level atom in a Hong-Ou-Mandel set-up. By computing the dynamics, we can describe the system at any time of the scattering event. This allows us to highlight the one-to-one correspondence between the saturation of the atom and the effective interaction induced between the photons. Furthermore, we discuss the integrability of the atomic beamsplitter and provide an intuitive picture for the correlations observed between the outgoing photons.
\end{abstract}


\maketitle

\section{Introduction}

The beamsplitter (BS) is an elementary unit of quantum linear optics~\cite{Bouland2014} and has applications in various fields, such as quantum computation~\cite{Knill2001,Kok2007} based on the Hong-Ou-Mandel (HOM) effect~\cite{HOM1987} or quantum network~\cite{Halder2007}. Recent experimental progress has brought the linear BS beyond the conventional optical realization made of glass, for instance using Landau-Zener transitions for electronic spin states in a double quantum dot~\cite{Petta2010} or for an artificial atom in a superconducting circuit~\cite{Oliver2005}, modulating SQUIDs in a superconducting cavity~\cite{DiVincenzo2010} or even using electromagnetically induced transparency for slow light in an atomic vapor cell~\cite{Xiao2008}.

\begin{figure}
\subfloat{
\includegraphics[width=0.12\textwidth]{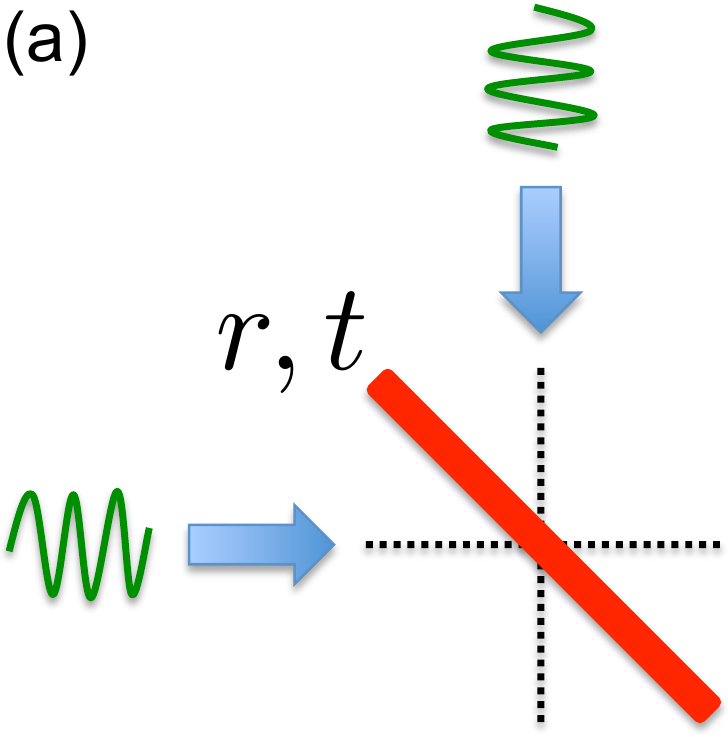}\label{fig:glassBS}
}\quad
\subfloat{
\includegraphics[width=0.31\textwidth]{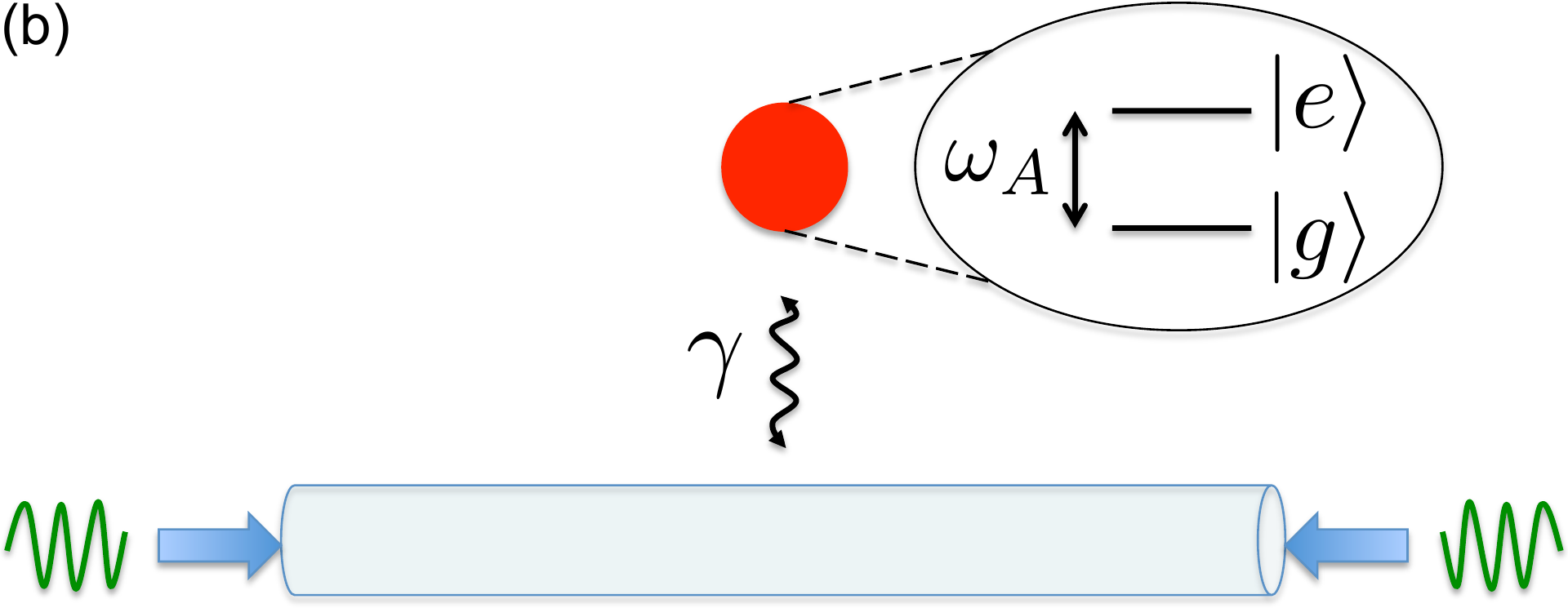}\label{fig:atomicBS}
}
\caption{\label{fig:BS}(color online).\quad Two indistinguishable photons impinging on a beamsplitter. (a) Conventional linear beamsplitter with reflection and transmission amplitudes respectively given by $r$ and $t$. (b) Atomic beamsplitter formed by strongly coupling a two-level atom to a 1D waveguide.}
\end{figure}

In the present work, we investigate the BS transformation realized by a two-level system (referred to as an “atom” in the rest of the paper) on photons propagating in a one-dimensional (1D) waveguide, as illustrated in Fig.\,\ref{fig:BS}. Strong light-matter interaction makes the 1D waveguide set-up~\cite{Babinec2010,Astafiev2010,Bleuse2011,Laucht2012,Arcari2014,Goban2014} a promising candidate for quantum information processing. Indeed, while photons are the preferred choice for communicating quantum information, their lack of interaction is a drawback for the implementation of two-qubit gates. In view of building an optical quantum computer, one way of introducing an effective interaction between photons at low light power is to use the atom as a nonlinear medium  \cite{Chang2014}. Indeed, because it cannot absorb or emit more than one photon at a time, a pair of incident photons will not interact with the atom in the same way as a single photon does. To date, several practical devices based on this nonlinearity have been proposed, such as single-photon transistors and routers~\cite{Chang2007,Bajcsy2009,Hoi2011}, on-chip amplifiers~\cite{Astafiev2010amp} or quantum non-demolition photodetectors~\cite{Witthaut2012}.

In fact, an effective two-qubit gate would also be possible in a regime of operation where the atomic BS behaves as a conventional linear BS. Specifically, when tuned as a balanced BS, two indistinguishable photons incoming in opposite directions will bunch into the same output mode. The output state thus reads $|\psi_{out}\rangle=\frac{|2,0\rangle+|0,2\rangle}{\sqrt{2}}$, corresponding to a superposition of finding the two photons co-propagating to the left or to the right in Fig\,\ref{fig:atomicBS}. This creation of entanglement between separated photons impinging on a BS \cite{Kim2002}, also called HOM effect, arises from the interference between probability amplitudes and is a direct manifestation of the bosonic nature of photons \cite{Jeltes2007}. It has been shown that it can be used to probabilistically realize quantum logic operations with only linear quantum optics and postselection on detection events \cite{Knill2001,Okamoto2011}. In this case, while the lack of photon-photon interaction precludes the realization of deterministic quantum computing \cite{Calsamiglia2002}, the atomic BS has the great advantage of being tunable \cite{SFopt2005}, in contrast to a piece of glass with fixed properties.

Following these motivations, we study analytically how the saturation of the atom effectively affects the behavior of the atomic BS in a HOM setup. Our transparent approach allows to \emph{compute the dynamics} and thus to intuitively understand our results by monitoring the atomic excitation during the scattering event. This differs from previous models limited to post-scattering descriptions based on a real-space formalism~\cite{SFSchwingerPRL2007,SFSchwinger2007,Zheng2010}, input-output theory~\cite{SFinout2010} or standard scattering theories \cite{Pletyukhov2012,Oehri2015}. Moreover, we study in details the correlations induced by the atom on the outgoing photons. While these correlations are highly non-trivial in the frequency domain \cite{SFSchwinger2007,Anders2015}, we provide an intuitive understanding of the underlying physics by studying them in the time domain. This lays the foundations for discussing the integrability of the device in various regimes of operation.

\section{Model}

We consider the system illustrated in Fig.\,\ref{fig:atomicBS}, consisting of a 1D waveguide strongly coupled to an atom with resonance frequency $\omega_A$ between ground $|g\rangle$ and excited $|e\rangle$ states. We study the scattering of two indistinguishable photons propagating in opposite directions \begin{equation}\label{statein}
|\psi_{in}\rangle = \int_{0}^\infty\! d \omega\,\int_{0}^\infty\! d \omega^\prime\, f(\omega)f(\omega^\prime)\,\hat{a}^\dagger_\omega \hat{b}^\dagger_{\omega^\prime} |0_a,0_b,g\rangle\ ,
\end{equation}
where $\hat{a}_\omega$ ($\hat{b}_\omega$) is the annihilation operator of the forward- (backward-) propagating photon mode and where the amplitude $f(\omega)$ is centered around frequency $\omega_0$ and has bandwidth $\Delta f\equiv\Omega$.

The dipole Hamiltonian describing the interaction between the atom and the propagating photons, under rotating wave approximation, is given by~\cite{Domokos2002} 
\begin{equation}\label{eq:Hamiltonian}
	\hat{H}_{dipole}=-i\hbar\int_{0}^\infty\!d \omega\, g_\omega \left[|e\rangle\langle g| \left(\hat{a}_\omega +\hat{b}_\omega \right)-\mathrm{H.c.} \right] ,
\end{equation}
where $g_\omega$ is the coupling constant. We shall work with the Weisskopf-Wigner approximation, so that the coupling enters the results through the atomic bandwidth $\gamma\equiv 2\pi g_{\omega_A}^2$.  Any relevant physical quantity is then readily obtained --- \emph{at any time t} --- by deriving the Heisenberg equation of the corresponding observable and solving a closed set of first-order differential equations (see Appendix \ref{sec:appA} for further details on the derivations).

The HOM effect is captured by the average coincidence after the scattering event, i.e. the probability of finding the two photons propagating in different output modes:
\begin{equation}\label{eq:Coindef}
	\mathcal{C}\equiv\lim\limits_{t\to\infty}\langle \psi_{in}|\hat{N}_a(t)\hat{N}_b(t)|\psi_{in}\rangle ,
\end{equation}
with $\hat{N}_a=\int_{0}^\infty\!d \omega\, \hat{a}_\omega^\dagger \hat{a}_\omega$ and $\hat{N}_b=\int_{0}^\infty\!d \omega\, \hat{b}_\omega^\dagger \hat{b}_\omega$ respectively the photon-number operators of forward and backward propagating modes. For a conventional 50/50 BS, the HOM bunching results in vanishing coincidence $\mathcal{C}=0$.

However, the average coincidence alone fails to capture the rich physics of the atomic BS. We expect this BS to be non-linear and frequency-dependent, and this means that the output photons may be distorted and correlated with each other; the extent up to which this happens may limit the integrability of the BS in a more complex circuit. For this more thorough study, we shall post-select on the cases when the outgoing photons exit in different modes (coincidence) and study the joint spectral distribution
\begin{equation}\label{eq:Sfreqdef}
	\mathcal{S}_{\omega_1,\omega_2}\equiv \lim\limits_{t\to\infty} |\langle 0_a,0_b,g|\hat{a}_{\omega_1}(t)\hat{b}_{\omega_2}(t)|\psi_{in}\rangle|^2\ .
\end{equation}
As it will turn out, some of the physics will be easier to understand from the two-time correlation
\begin{equation}\label{eq:Stimedef}
	\mathcal{S}_{\tau_1,\tau_2}(t)\equiv |\langle 0_a,0_b,g|\hat{a}_{\tau_1}(t)\hat{b}_{\tau_2}(t)|\psi_{in}\rangle|^2\ ,
\end{equation}
where $\hat{a}_{\tau_1}(t)\equiv\int_{0}^\infty\!d \omega\, \hat{a}_{\omega}(t)e^{-i \omega \tau_1}$ and similarly for $\hat{b}_{\tau_2}(t)$.

Finally, by tracking the atomic excitation during the scattering event, we can correlate the saturation of the atom and the induced nonlinearity on the photons. We shall therefore also study the probability of excitation of the atom as a function of time $t$
\begin{equation}\label{eq:Exdef}
	\mathcal{P}_e(t)\equiv \frac{\langle \psi_{in}|\hat{\sigma}_z(t)|\psi_{in}\rangle+1}{2} ,
\end{equation}
where $\hat{\sigma}_z=|e\rangle\langle e|-|g\rangle\langle g|$. $\mathcal{P}_e(t)$ varies between 0, corresponding to the atom being in the ground state $|g\rangle$, and 1 when the atom is in the excited state $|e\rangle$.

\section{Monochromatic light}

\begin{figure}
\includegraphics[width=0.46\textwidth]{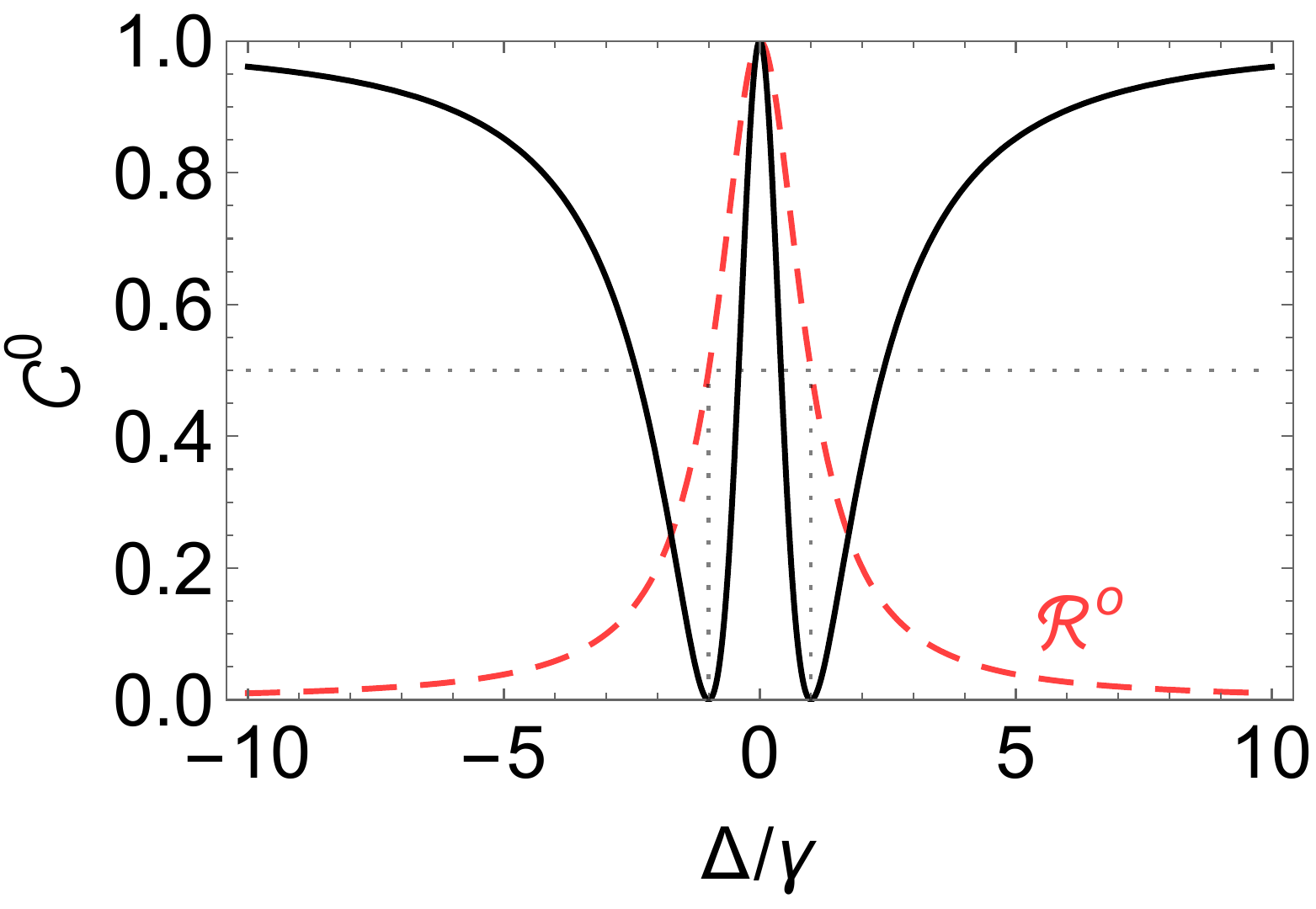}
\caption{\label{fig:Cmono}(color online).\quad Average coincidence $\mathcal{C}^{o}$ \eqref{eq:Cmono} at the output of the atomic BS as a function of the normalized detuning $\Delta/\gamma$ for monochromatic indistinguishable photons. In dashed red is the reflection coefficient $\mathcal{R}^{o}$ for a monochromatic single-photon input. The behaviour is that of a linear, frequency-independent BS. At resonance, $\mathcal{R}^{o}=1$ (as observed in recent experiments~\cite{Astafiev2010,Hoi2011}) and therefore each photon is reflected. When $|\Delta|=\gamma$, $\mathcal{R}^{o}=1/2$ and HOM bunching is predicted.}
\end{figure}

We first investigate the monochromatic limit $\Omega\ll\gamma$, frequently used in theoretical studies~\cite{SFSchwingerPRL2007,Fan2013}. The details of the calculations are given in Appendix \ref{sec:appA}. The average coincidence probability is found to be (Fig.~\ref{fig:Cmono})
\begin{equation}\label{eq:Cmono}
	\mathcal{C}^{o}\approx 1 - \frac{4 (\Delta/\gamma)^2}{[1 + (\Delta/\gamma)^2]^2} = 1-4 \mathcal{R}^{o}\mathcal{T}^{o} ,
\end{equation}
with $\Delta\equiv\omega_0-\omega_A$ the detuning and $\mathcal{R}^{o}=1-\mathcal{T}^{o}=1/[1+(\Delta/\gamma)^2]$ the single-photon reflection coefficient in the monochromatic limit~\cite{SFopt2005}. In other words, for monochromatic photons, the average two-photon coincidence $\mathcal{C}^{o}$ is fully determined by the atomic response to individual single photons and follows the behavior of a linear BS with reflection coefficient $\mathcal{R}^{o}$. This means that \emph{no interaction between the photons is mediated by the atom.} This absence of nonlinearity correlates well with the fact that
\begin{equation}\label{eq:Exlin}
	\mathcal{P}_e^{o}(t) \approx 0 \quad \forall t
\end{equation}
meaning that the atom is mostly found in the ground state during the scattering event. This so-called weak-excitation limit arises naturally: the narrow frequency bandwidth of the incoming photons $\Omega\ll\gamma$ implies a long pulse in the time domain compared to the atomic lifetime.

Finally we find that the spectral distribution of the outgoing photons is preserved:
\begin{equation}
	\mathcal{S}^{o}_{\omega_1,\omega_2}/\mathcal{C}\approx |f(\omega_1) f(\omega_2)|^2 \,.
\end{equation} This is again explained by the monochromatic limit: since the atom effectively sees a single frequency $\omega_0$, it responds in the same way to each frequency components of the input pulse. Thus, \emph{no correlation between the photons is induced by the atom.}

\begin{figure}
\includegraphics[width=0.46\textwidth]{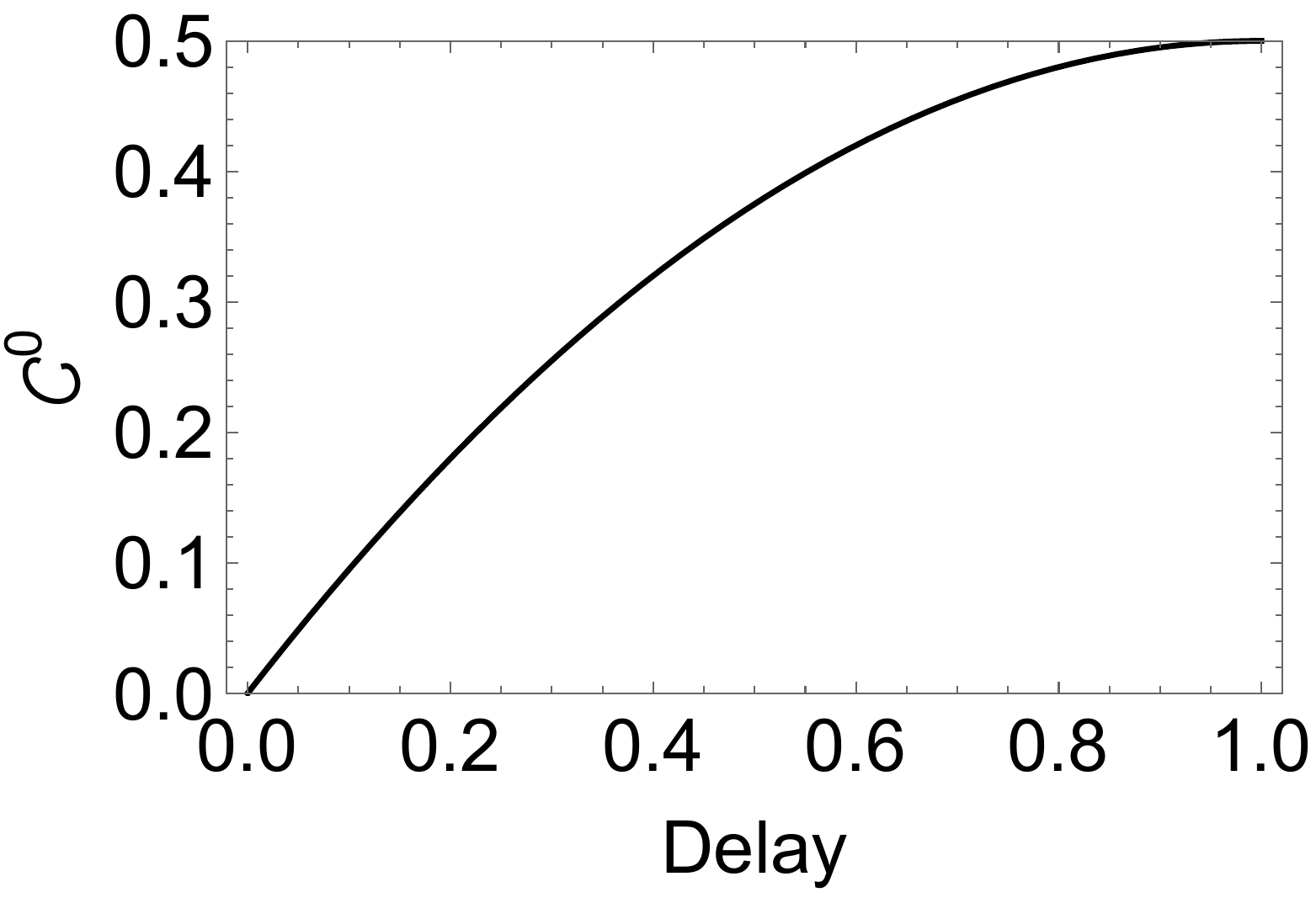}
\caption{\label{fig:CmonoHOMdip}Average coincidence at the output of the balanced atomic BS as a function of the delay between the two photons in the effectively monochromatic regime. The delay is normalized with respect to the photon-pulse duration and reaches 1 when the pulses do not overlap anymore.}
\end{figure}

As a last check, we can study how the coincidence varies when the two input photons are not arriving simultaneously at the BS. To this effect, we replace $f(\omega)f(\omega')\longrightarrow f(\omega)f(\omega^\prime)e^{i\omega^\prime\tau}$ in \eqref{statein}. Fig.~\ref{fig:CmonoHOMdip} shows the result at the point $\mathcal{R}^{o}=\mathcal{T}^{o}=1/2$: as expected, the HOM bunching at $\tau=0$ disappears when the delay is large enough for the photons become distinguishable.

\section{Pulsed light: frequency dependence and nonlinear regime}

In summary, we have seen that for $\Omega\ll\gamma$ the atomic BS is both \textit{linear} and \textit{photon-shape preserving}. It is important to notice that these properties are not identical: distortion and correlation of the photon shapes are expected as soon as the frequency dependence of the BS plays a role. In order to clearly separate these effects from actual non-linearities, we discuss first the behaviour of a hypothetical BS that would be linear but frequency-dependent (section \ref{sec:linBS}); only later we'll compute the physics of the actual atomic BS (section \ref{sec:nlBS}). For the remainder of the paper we focus on the case when the photons are at resonance, $\Delta=\omega_0-\omega_A=0$.

\subsection{Linear BS with frequency-dependent response}\label{sec:linBS}

Consider a linear BS whose reflection $r_\omega$ and transmission $t_\omega$ amplitudes are frequency-dependent:
\begin{eqnarray}
	\left\{\begin{aligned}\hat{a}^\dagger_\omega&\to& t_\omega \hat{a}^\dagger_\omega+r_\omega \hat{b}^\dagger_\omega\\
	\hat{b}^\dagger_\omega&\to& t_\omega \hat{b}^\dagger_\omega+r_\omega \hat{a}^\dagger_\omega \end{aligned}\right. 
\end{eqnarray}
The input state \eqref{statein} is then mapped to
\begin{widetext}
\begin{equation}\label{eq:outStateLin}
	|\psi_{out}\rangle=\int_{0}^\infty\! d \omega\,\int_{0}^\infty\! d \omega^\prime\, f(\omega) f(\omega^\prime) \Big[(t_\omega t_{\omega^\prime}+r_\omega r_{\omega^\prime}) \hat{a}^\dagger_\omega \hat{b}^\dagger_{\omega^\prime}+t_\omega r_{\omega^\prime}(\hat{a}^\dagger_\omega \hat{a}^\dagger_{\omega^\prime}+\hat{b}^\dagger_\omega \hat{b}^\dagger_{\omega^\prime})\Big]|0_a,0_b,g\rangle ,
\end{equation}
\end{widetext}
from which we can compute the coincidence as \cite{Anders2015}
\begin{eqnarray}\label{eqApp:linearC}
	\mathcal{C}&=&\int_{0}^\infty\! d \omega\,\int_{0}^\infty\! d \omega^\prime\, |f(\omega) f(\omega^\prime) (t_\omega t_{\omega^\prime}+r_\omega r_{\omega^\prime})|^2\nonumber\\
	&=& \mathcal{T}^2+\mathcal{R}^2+2\operatorname{Re}\Big[(\int_{0}^\infty\! d \omega\, |f(\omega)|^2 t_\omega r_{\omega}^*)^2\Big] .
\end{eqnarray}
where $\mathcal{R}=1-\mathcal{T}=\int_{0}^\infty\! d \omega\, |f(\omega) r_\omega|^2$ is the single-photon reflection coefficient for the pulse under consideration.

Assume now that $(r_\omega,t_{\omega})$ are those of a single photon impinging on the atomic BS \cite{SFopt2005}
\begin{equation}
		r_\omega=\frac{-i}{(\omega-\omega_A)/\gamma+i}\quad\text{and}\quad
		t_\omega=\frac{(\omega-\omega_A)/\gamma}{(\omega-\omega_A)/\gamma+i}\nonumber\,.
\end{equation}
Then, the product $t_\omega r_\omega^*=i\frac{(\omega-\omega_A)/\gamma}{[(\omega-\omega_A)/\gamma]^2+1}$ is an odd function in $\omega-\omega_A$. Therefore, any symmetric pulse of finite bandwidth on resonance with the atomic transition would yield in the linear regime
\begin{equation}\label{eq:nonmonolinRes}
	\mathcal{C}=\mathcal{T}^2+\mathcal{R}^2=1-2\mathcal{R}\mathcal{T}\geq\frac{1}{2}\ ,
\end{equation}
because the last term in \eqref{eqApp:linearC} vanishes due to the parity of $|f(\omega-\omega_A)|^2$ \footnote{Note that Eq.~\eqref{eq:nonmonolinRes} is only valid at resonance $\mathcal{C}(\Omega,\Delta=0)=1-2\mathcal{R}(\Omega)\mathcal{T}(\Omega)$. In the limit of vanishingly small bandwidth, where resonant photons are fully reflected, it agrees with the unit coincidence \eqref{eq:Cmono} obtained in the monochromatic regime $\mathcal{C}(\Omega\ll\gamma,\Delta=0)=1-2\mathcal{R}(\Omega\ll\gamma)\mathcal{T}(\Omega\ll\gamma)=1=1-4\mathcal{R}^o(\Delta=0)\mathcal{T}^o(\Delta=0)$.}.

The spectral distribution of the outgoing photons follows from Eq.~(\ref{eq:outStateLin}) as
\begin{equation}\label{eq:linearSpec}
	\mathcal{S}_{\omega_1,\omega_2}=|f(\omega_1) f(\omega_2) (t_{\omega_1} t_{\omega_2}+r_{\omega_1} r_{\omega_2})|^2\ .
\end{equation}
As expected in the linear regime, it is a simple interference between the two photons being either transmitted or reflected. From this decomposition, one can see that frequencies satisfying
\begin{equation}\label{eq:linDest}
	t_{\omega_1} t_{\omega_2}=-r_{\omega_1} r_{\omega_2}\implies (\omega_1-\omega_A)(\omega_2-\omega_A)=\gamma^2
\end{equation} interfere destructively. This naturally includes the situation where both frequencies are the same and equal to $\omega_1=\omega_2=\omega_A+\gamma$: the physics of the linear HOM effect is still present, as it should. Besides, the presence of a continuum of frequencies in the input pulses allows for more combinations which also lead to a vanishing coincidence. However, it also gives rise to many other combinations that contribute significantly to the coincidence. Frequency components close to resonance and far off-resonant are respectively reflected and transmitted without interfering; and those frequencies that satisfy
\begin{equation}\label{eq:linConst}
	t_{\omega_1} t_{\omega_2}=r_{\omega_1} r_{\omega_2}\implies (\omega_1-\omega_A)(\omega_2-\omega_A)=-\gamma^2\ .
\end{equation} produce constructive interference. All together, these contributions lead to the lower bound on the coincidence in Eq.~\eqref{eq:nonmonolinRes}.

\subsection{Nonlinearity induced by the atomic BS}\label{sec:nlBS}

We can now study the exact response of the atomic BS to pulses of finite bandwidth $\Omega$. For simplicity of the formulas and the presentation of the results, we work with square pulses unless specified otherwise.

\begin{figure}
\includegraphics[width=0.46\textwidth]{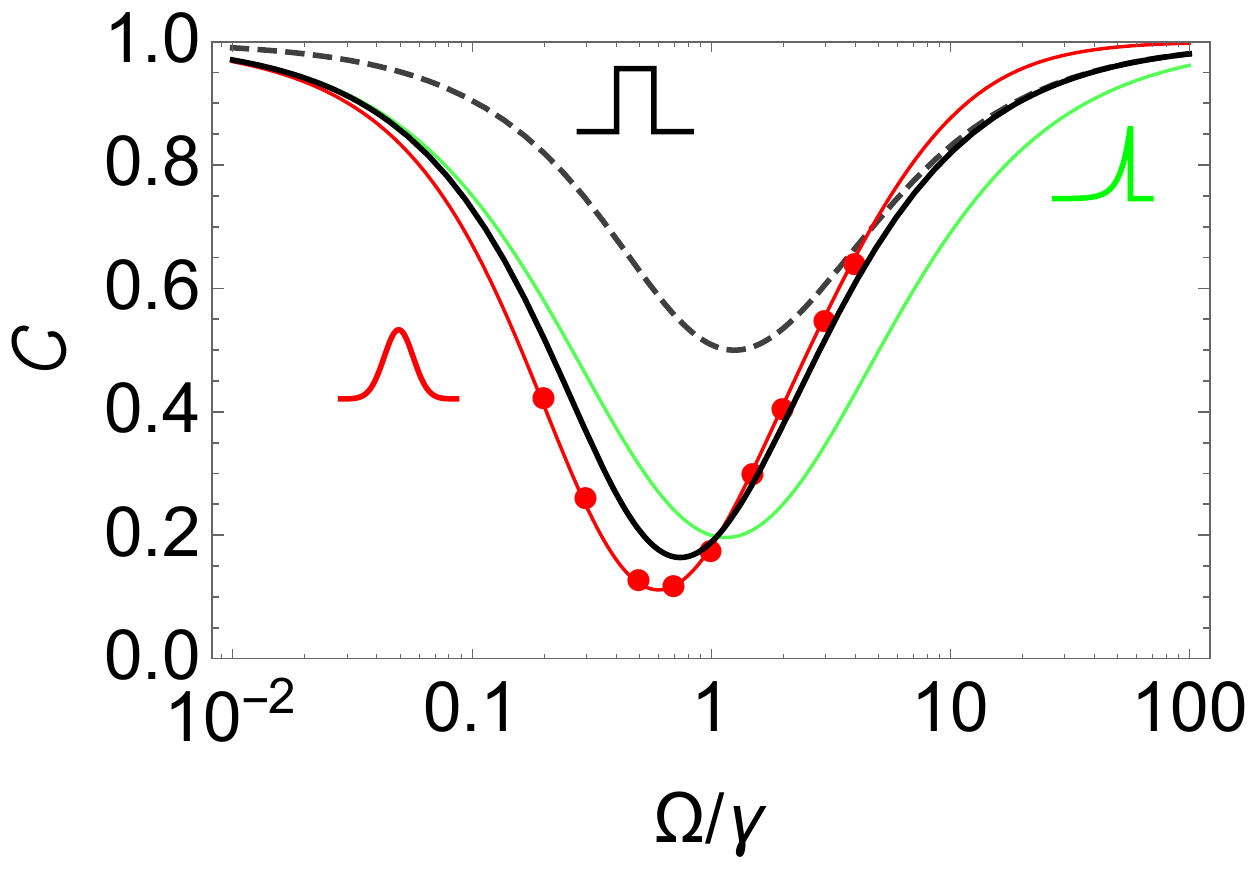}
\caption{\label{fig:Cnonlin}(color online).\quad Average coincidence at the output of the BS on resonance $\Delta=0$ as a function of the normalized bandwidth $\Omega/\gamma$. The thick line is the coincidence derived for the atomic BS (\ref{eq:Cnonlin}). The dashed line represents the expected coincidence $1-2 \mathcal{R}^{\sqcap}\mathcal{T}^{\sqcap}$ for square pulses impinging on a linear BS with reflection coefficient $\mathcal{R}^{\sqcap}$. The red filled circles are numerical simulations obtained in Ref.\,\cite{Anders2014} for gaussian pulses and the red line is computed with our model. The green line corresponds to exponentially rising pulses.}
\end{figure}

Let us start with the average coincidence, which is found to be (see Appendix \ref{sec:appA})
\begin{eqnarray}\label{eq:Cnonlin}
	\mathcal{C}^{\sqcap}&=&1 - 3\Omega/\gamma\left[1-\Omega/\gamma+e^{-\frac{2}{\Omega/\gamma}}(1+\Omega/\gamma)\right]\end{eqnarray}
where $\mathcal{R}^{\sqcap}=1-\mathcal{T}^{\sqcap}=1+(-1+e^{-\frac{2}{\Omega/\gamma}})\frac{\Omega/\gamma}{2}$ is the single-photon reflection coefficient for a square pulse of bandwidth $\Omega$. In particular, $\mathcal{C}^{\sqcap}$ is different from $1-2 \mathcal{R}^{\sqcap}\mathcal{T}^{\sqcap}$ the prediction of the linear BS. As we see in Fig.\,\ref{fig:Cnonlin}, the effect of non-linearity is maximal when $\Omega\approx\gamma$. For instance, when the BS is balanced at the single-photon level ($\mathcal{R}^{\sqcap}=\mathcal{T}^{\sqcap}=1/2$), which happens for $\Omega/\gamma\approx1.25$, Eq.~\eqref{eq:nonmonolinRes} for the linear BS yields $\mathcal{C}=1/2$, while Eq.~(\ref{eq:Cnonlin}) predicts $\mathcal{C}^{\sqcap}\approx 0.23$ for the atomic BS.

To leave it clear that our method is not limited to square pulses, in Fig.~\ref{fig:Cnonlin} we have also plotted $\mathcal{C}$ for exponentially rising pulses (time-reversed of atomic spontaneous emission) and for Gaussian pulses. The latter are found to be in perfect agreement with the recent numerical simulations performed by A. Nysteen \textit{et al.}~\cite{[{}][{ The data plotted here corresponds to Fig.\,6 of the paper and was kindly shared by the authors.}]Anders2014} on the same physical situation. In all these cases, non-linear effects show a violation of the linear lower bound \eqref{eq:nonmonolinRes}.

\begin{figure}
\includegraphics[width=0.46\textwidth]{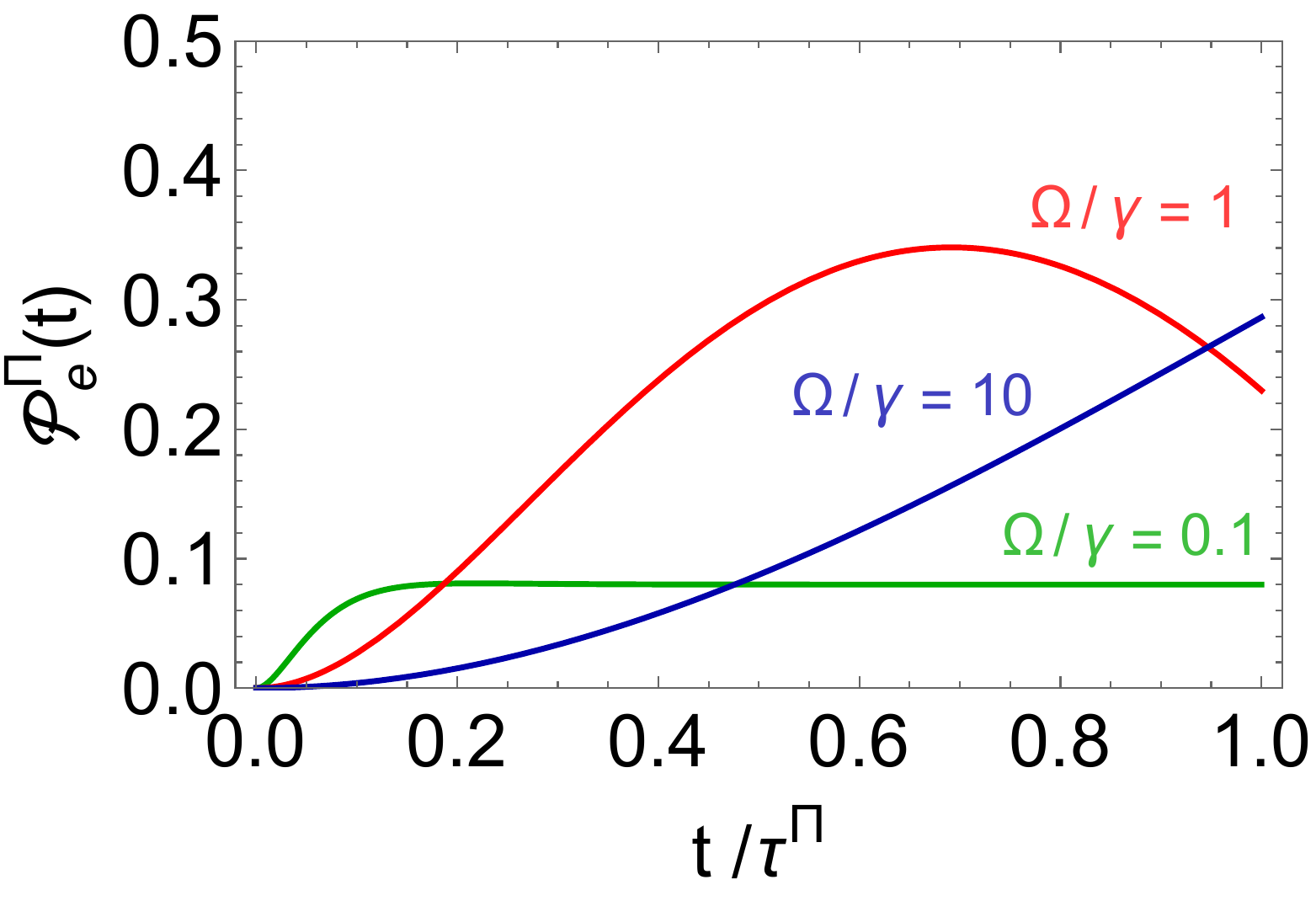}
\caption{\label{fig:Exnonlin}(color online).\quad Probability of atomic excitation as a function of time $t$ in units of pulse duration $\tau^\sqcap=2/\Omega$. The green, red and blue line represent respectively the normalized bandwidth $\Omega/\gamma=0.1$, $\Omega/\gamma=1.25$ and $\Omega/\gamma=10$. $\Delta=0$.}
\end{figure}

As before, we expect to find a correlation between the nonlinear response of the atomic BS and the atomic excitation during the scattering event. This can be seen by evaluating the atomic excitation during the scattering event (incidentally, a time-dependent quantity that is not accessible to post-scattering descriptions \cite{SFSchwingerPRL2007}). We find (see Appendix \ref{sec:appA})
\begin{eqnarray}\label{eq:Exnonlin}
	\mathcal{P}_e^{\sqcap}(t)&=&\Omega/\gamma\Big[1-2\Omega/\gamma+2e^{-\gamma t}\Big(-1+(-1+\gamma t)4\Omega/\gamma\Big)\nonumber\\
	&& +e^{-2\gamma t} \Big(1+(5+2\gamma t)2\Omega/\gamma\Big)\Big] .
\end{eqnarray}
As shown in Fig.\,\ref{fig:Exnonlin}, when $\Omega\approx\gamma$ the atom is significantly excited during most of the pulse duration, and it's in this regime that the largest nonlinearity is present. When we move away from the regime $\Omega\approx\gamma$, the nonlinearity induced by the atom decreases. For $\Omega\ll\gamma$, the atom is only weakly excited, as we discussed above (\ref{eq:Exlin}). For $\Omega\gg\gamma$, the excitation builds up slowly; in the frequency domain, most of the components of the pulse are off-resonant. In summary, the regime $\Omega\approx\gamma$ thus combines the advantages of pulses concentrated in time (higher intensity) and in frequency (resonant coupling), leading to significant nonlinearity induced by the atom. This corroborates conclusions reached in a cavity-based system~\cite{Rosenblum} and is also in agreement with findings obtained for coherent states~\cite{Chang2007}.

\begin{figure*}
\begin{minipage}{0.9\textwidth}
\subfloat{
\includegraphics[width=0.31\textwidth]{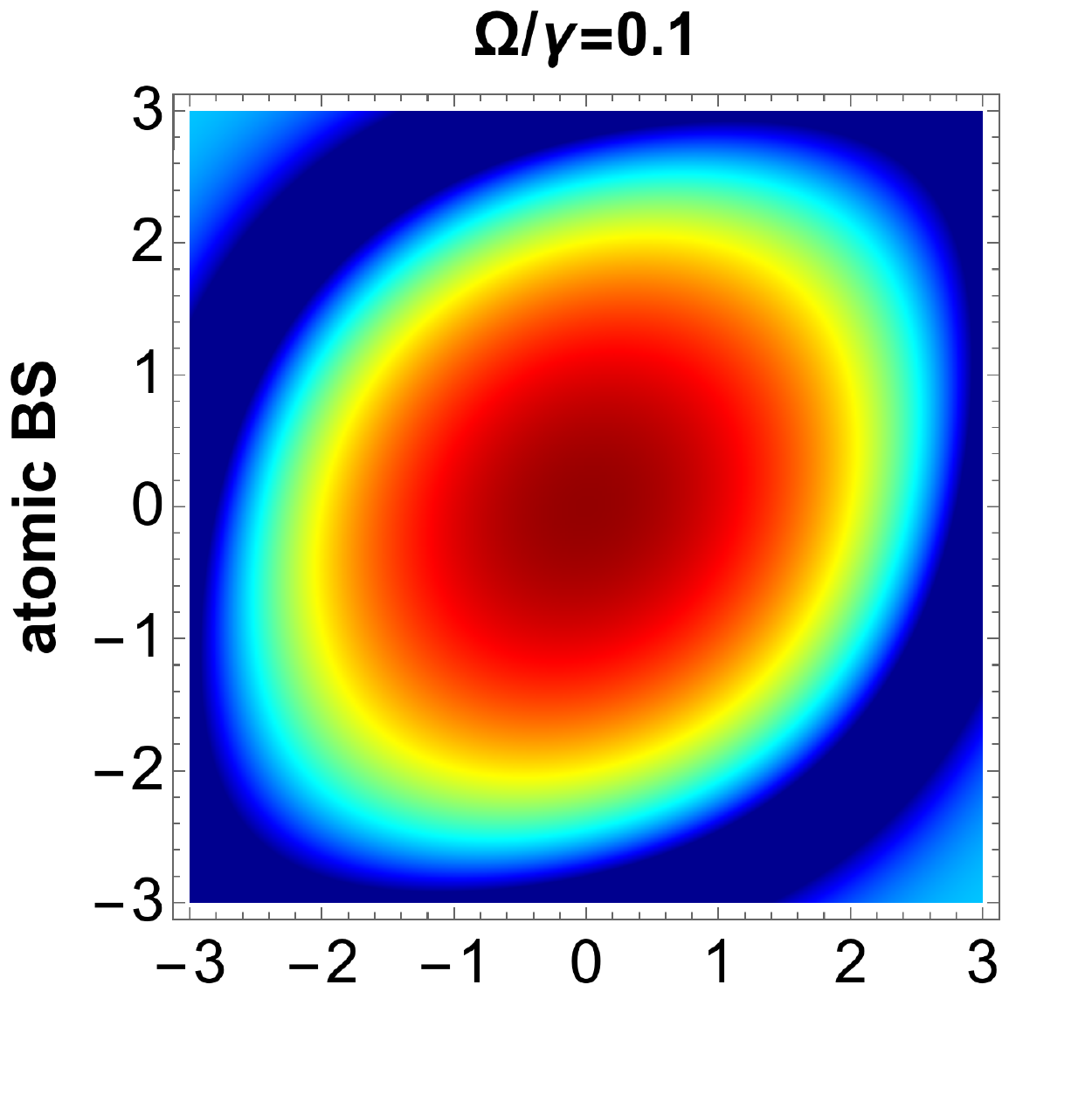}
}
\subfloat{
\includegraphics[width=0.31\textwidth]{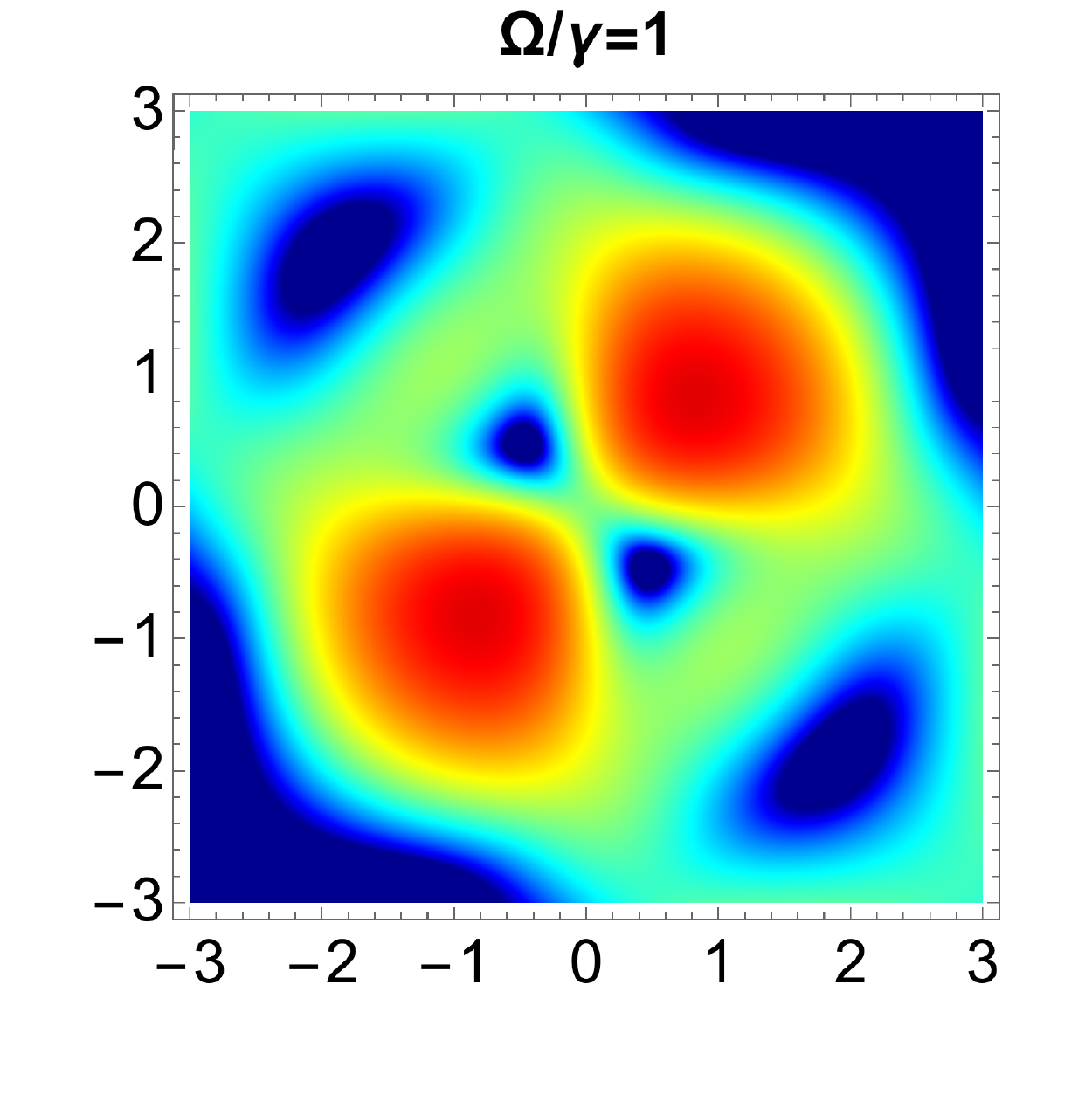}
}
\subfloat{
\includegraphics[width=0.31\textwidth]{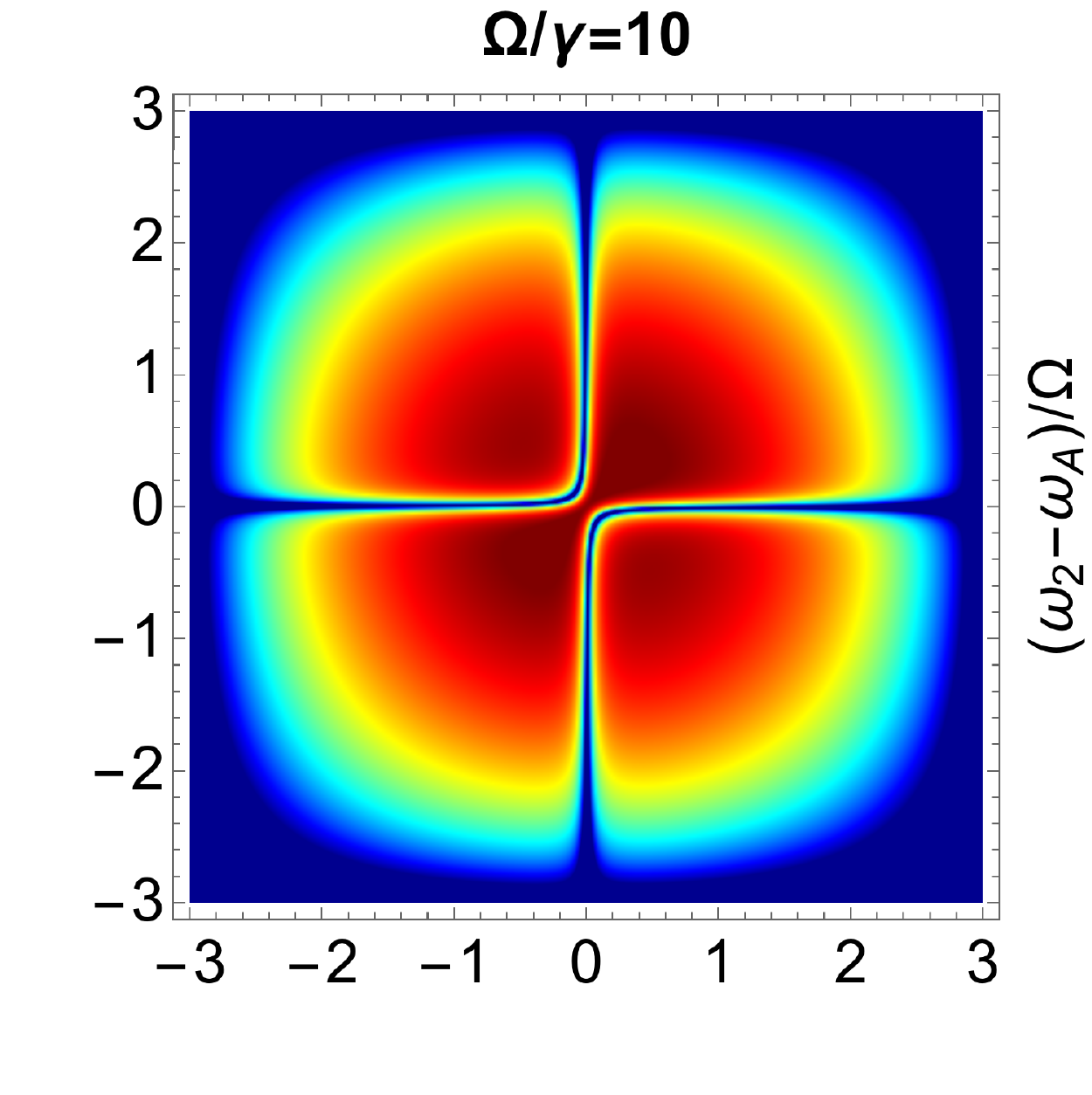}
}\\\vspace{-0.5cm}
\subfloat{
\includegraphics[width=0.31\textwidth]{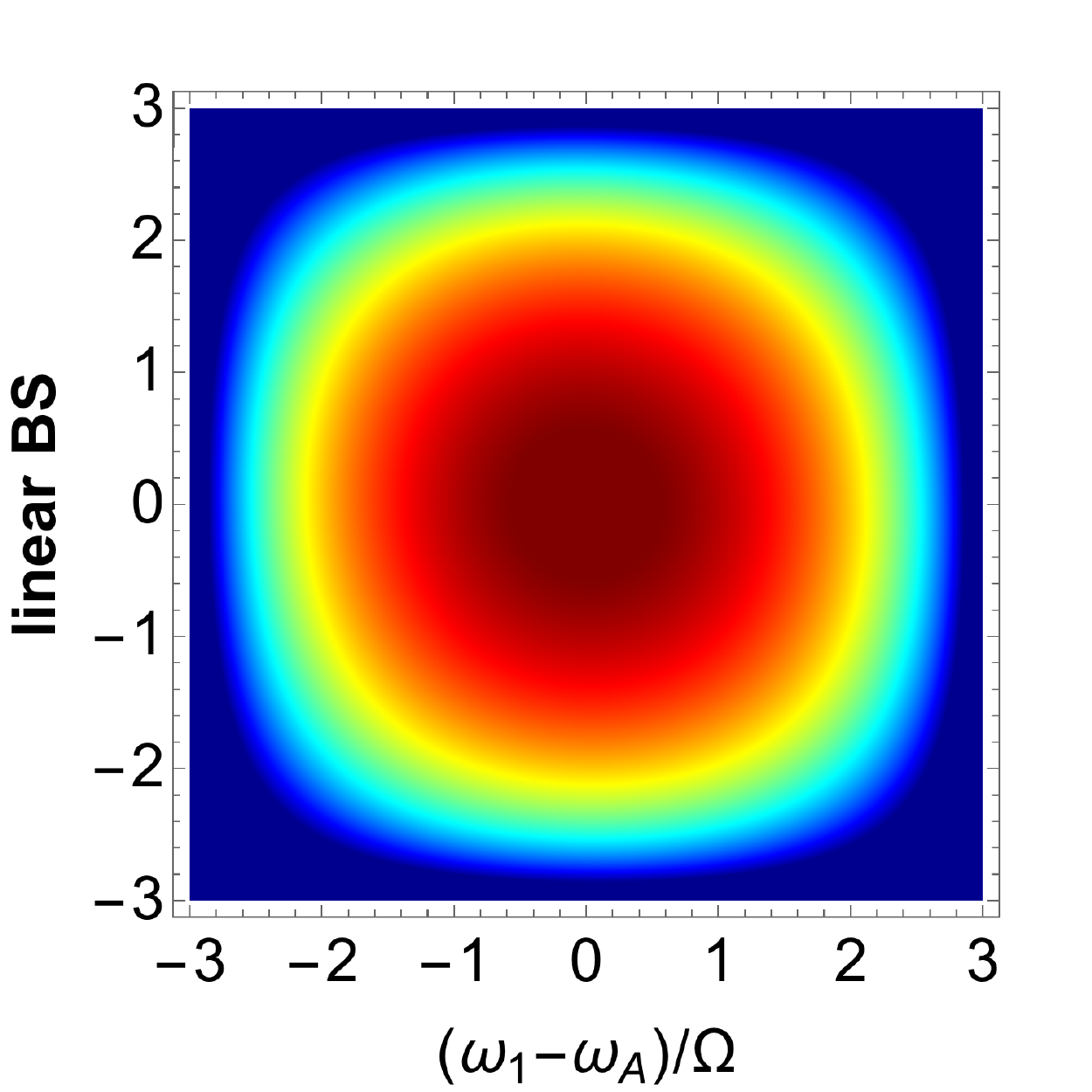}
}
\subfloat{
\includegraphics[width=0.31\textwidth]{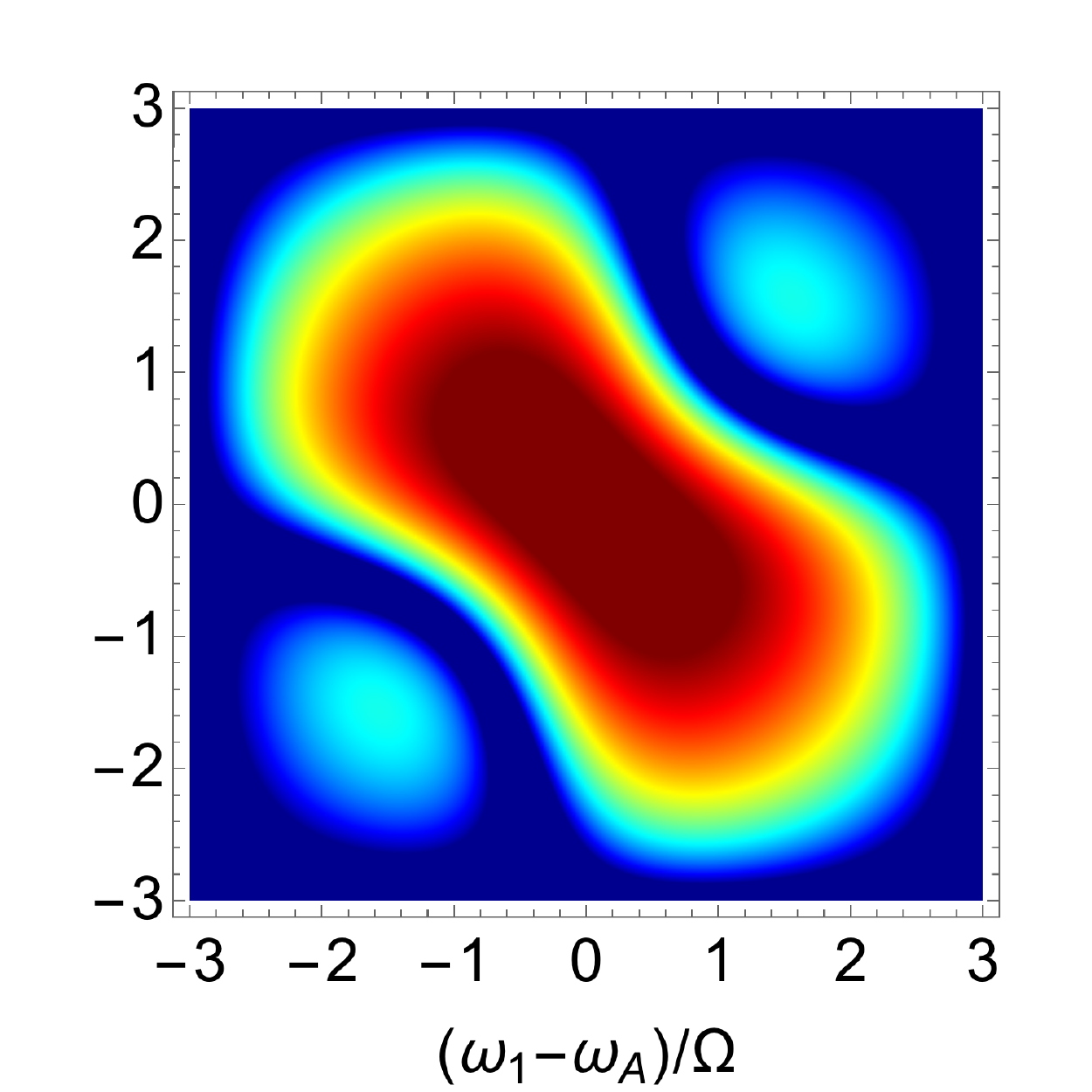}
}
\subfloat{
\includegraphics[width=0.31\textwidth]{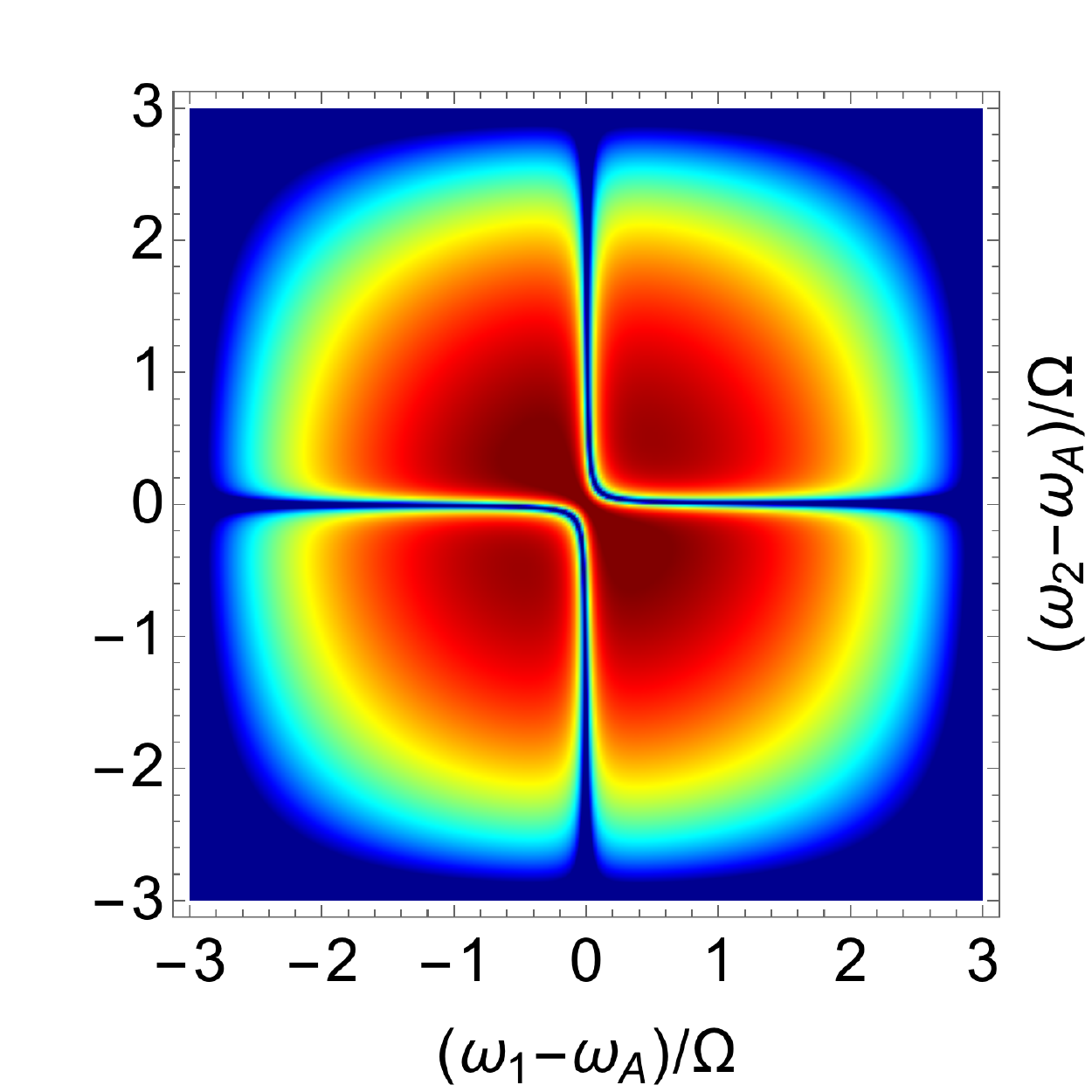}
}
\end{minipage}\begin{minipage}{0.1\textwidth}
\subfloat{
\quad\includegraphics[width=0.8\textwidth]{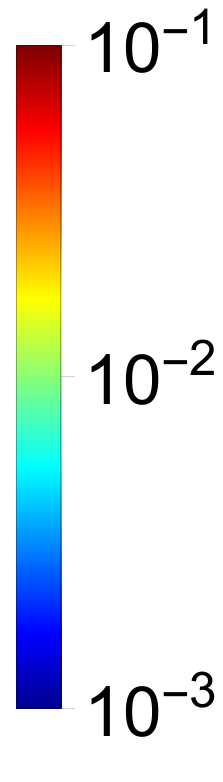}
}
\end{minipage}
\caption{\label{fig:freqDensity}(color online).\quad The spectral distribution of the outgoing photons post-selected on coincidence events $\mathcal{S}^{\sqcap}_{\omega_1,\omega_2}/\mathcal{C}$ for resonant square pulses of various bandwidth $\Omega$. The first row corresponds to the photons being scattered by the atomic BS. The second row, meant for comparison, is the fictitious situation where the scatterer would be a linear BS with the same reflection $r_\omega$ and transmission $t_\omega$ as those of the atomic BS.}
\end{figure*}

\subsection{Correlations in the frequency domain}

While the emergence of the nonlinearity in a specific range of bandwidth can be correlated to the saturation of the atom, the reason for the resulting increased bunching does not appear as straightforward. Let us turn first to the correlations of the outgoing photons in the frequency domain, which read (see Appendix \ref{sec:appB})
\begin{eqnarray}\label{eq:nonlinearSpec}
	&&\mathcal{S}_{\omega_1,\omega_2}^{\sqcap}=\big|f(\omega_1) f(\omega_2) (t_{\omega_1} t_{\omega_2}+r_{\omega_1} r_{\omega_2})\\
	&&+\frac{r_{\omega_1}+r_{\omega_2}}{\pi\gamma}\int_{0}^\infty\!d \omega\,f(\omega)f(\omega_1+\omega_2-\omega)r_{\omega}r_{\omega_1+\omega_2-\omega}\big|^2 \ .\nonumber
\end{eqnarray}
We recognize here the sum of two amplitudes, the linear term obtained in Eq.~(\ref{eq:linearSpec}) and of what has been called ``background fluorescence'' in Ref.~\cite{SFSchwingerPRL2007}. There, it is understood as a redistribution of the input photons frequencies which ``arises as one photon inelastically scatters off a composite transient object formed by the atom absorbing the other photon.'' This is definitely a qualitative description of the role of that term, but not of its details.

More insight can be gathered from Fig.~\ref{fig:freqDensity}, where we plot $\mathcal{S}_{\omega_1,\omega_2}^{\sqcap}/\mathcal{C}$ for square pulses of various bandwidth $\Omega$, together with the linear term alone. The case of small bandwidth $\Omega/\gamma=0.1$ is expected: in the monochromatic regime, the atomic BS was found to be linear and shape preserving. When moving away from the monochromatic regime, the linear term shows distinctive destructive interference for frequencies satisfying Eq.~(\ref{eq:linDest}). The actual response of the atomic BS, however, is far more complex. We would like to highlight one feature, that we'd call \emph{reversed HOM effect}: destructive interference are observed for the exact frequency combinations which were yielding constructive interference in the linear regime and \emph{vice versa}.

\def\imagetop#1{\vtop{\null\hbox{#1}}}
\begin{table*}
\caption{\label{tab:pathAmp} All the paths interfering to yield a coincidence event. Their respective amplitude is given for the case of a linear BS in the frequency and time domain. (a) Both photons passed by the atom without interacting. (b) One of the photons has been absorbed and reemitted in the forward direction while the other did not interact. (c) Both photons have been absorbed and reemitted in opposite directions. The non-linearity of the atomic BS is going to modify the amplitudes of paths (c) according to Eqs \eqref{totalcorrect} and \eqref{eq:nlCorrect}.}
\begin{ruledtabular}
\begin{tabular}{cccccc}\vspace{-0.2cm}
 & (a) & \multicolumn{2}{c}{(b)} & \multicolumn{2}{c}{(c)} \\\vspace{0.1cm}
Path & \imagetop{\includegraphics[width=0.1\textwidth]{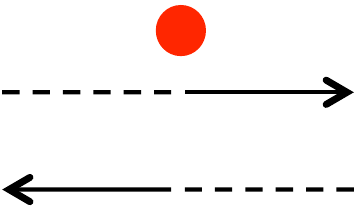}}\quad & \imagetop{\includegraphics[width=0.1\textwidth]{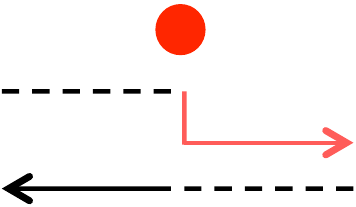}}\quad & \imagetop{\includegraphics[width=0.1\textwidth]{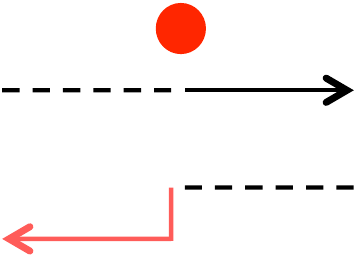}}\quad & \imagetop{\includegraphics[width=0.1\textwidth]{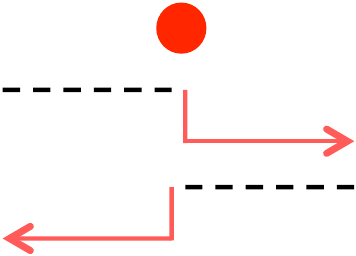}} & \imagetop{\includegraphics[width=0.105\textwidth]{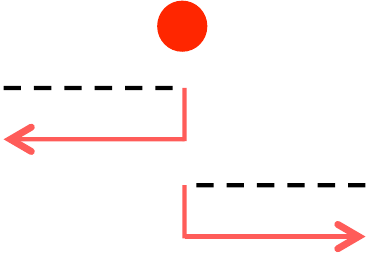}} \\
\hline
Frequency & $f(\omega_1)f(\omega_2)$ & $f(\omega_1)f(\omega_2)r_{\omega_1}$ & $f(\omega_1)f(\omega_2)r_{\omega_2}$ &  $f(\omega_1)f(\omega_2)r_{\omega_1}r_{\omega_2}$ & $f(\omega_1)f(\omega_2)r_{\omega_1}r_{\omega_2}$  \\
Time & $\xi(\tau_1)\xi(\tau_2)$ &  $c_A(\tau_1)\xi(\tau_2)$ & $\xi(\tau_1)c_A(\tau_2)$ & $c_A(\tau_1)c_A(\tau_2)$ & $c_A(\tau_1)c_A(\tau_2)$ \\
\end{tabular}
\end{ruledtabular}
\end{table*}

Instead of trying to further infer the physics of the atomic response from the post-scattering correlations in frequency domain, let's tackle the problem at the source, breaking down the different path amplitudes contributing to a coincidence event. This is represented in Table~\ref{tab:pathAmp}, where the amplitudes are given in the case of a linear BS. One can check that by summing the five paths we recover the linear spectra obtained in Eq.~(\ref{eq:linearSpec}). Now we also understand that the saturation of the atom will affect the (c) paths, those in which both photons are absorbed and subsequently re-emitted. In particular, the nonlinear contribution is expected to correct the amplitude of these two paths in a way that accounts for the impossibility of the atom to emit two photons at the same time. Since this is not obvious in the frequency domain (see Eq.~(\ref{eq:nonlinearSpec})), we continue the study in the time domain, where the typical response time of the atom $\gamma^{-1}$ should appear naturally.

\subsection{Correlations in the time domain}

In the time domain, instead of asking for the probability density of detecting a photon at a given frequency $\omega$ (Eq.~\ref{eq:Sfreqdef}), we ask for the probability density of detecting it at time $\tau$ (Eq.~\ref{eq:Stimedef}). This parameter $\tau$, which has to be distinguished from the dynamical time $t$, represents the time distance from the wavefront of the pulse.

The time decomposition of the input pulse is readily obtained as a Fourier transform of the frequency shape $\xi(\tau)\equiv \mathcal{F}[f(\omega)](\tau)$. This accounts for the process in Table~\ref{tab:pathAmp}(a). Less trivially, the linear amplitude of detecting a photon that has been absorbed and reemitted by the atom at a distance $\tau$ from the wavefront is given by the convolution between the characteristic response function of the atom and the incoming pulse profile:
\begin{equation}\label{eq:linAmp}
	c_A(\tau)\equiv \mathcal{F}[f(\omega)r_\omega](\tau) =-\gamma\int_{0}^\tau\!\mathrm{d} \tau^\prime\, e^{-\gamma(\tau-\tau^\prime)}\xi(\tau^\prime)\ .
\end{equation}
The convolution is run from the wavefront until the time distance of interest $\tau$, which accounts for the fact that this photon could have been absorbed at any moment right from the start of the pulse. Given the symmetric coupling of the atom to the waveguide, this amplitude does not depend on the direction of emission.

With this knowledge we can now move on to describe the full correlations in the time domain, which read (see Appendix \ref{sec:appB})
\begin{eqnarray}\label{eq:nonlinearTime}
	&&\mathcal{S}_{\tau_1,\tau_2}^{\sqcap}(t)=\big|\xi(\tau_1)\xi(\tau_2)+[\theta(t-\tau_2)\times\\
	&&(\xi(\tau_1)c_A(\tau_2)+\theta(\tau_2-\tau_1)2c_A(\tau_1)c_A^\text{nl}(\tau_2,\tau_1))+1\leftrightarrow 2]\big|^2\ ,\nonumber
\end{eqnarray}
where the Heaviside functions that put conditions on the dynamical time $t$ translate the fact that one cannot observe a photon emitted by the atom in the part of the pulse that has not reached it yet. When comparing this result with the linear correlations shown in Table~\ref{tab:pathAmp}, it appears that the amplitude of the paths (c) -- which involve the absorption and emission of both photons -- has been mapped to
\begin{equation}\label{totalcorrect}
	c_A(\tau_1)c_A(\tau_2)\to \theta(\tau_2-\tau_1)c_A(\tau_1)c_A^\text{nl}(\tau_2,\tau_1)+1\leftrightarrow 2\ ,
\end{equation}
with a nonlinear amplitude
\begin{equation}\label{eq:nlCorrect}
	c_A^\text{nl}(\tau_2,\tau_1)=-\gamma\int_{\tau_1}^{\tau_2}\!\mathrm{d} \tau^\prime\, e^{-\gamma(\tau_2-\tau^\prime)}\xi(\tau^\prime)\ ,
\end{equation}
which is formally identical to Eq.~(\ref{eq:linAmp}) but with the starting point of the integration shifted away from the wavefront. This result is easily interpreted as follows: if the first photon is emitted at a distance $\tau_1$ and the second at $\tau_2$, then the second photon can only have been absorbed during the time interval $\tau_2-\tau_1$. In other words, \emph{the second photon had less time to be absorbed than if it had been interacting alone with the atomic BS}. Therefore the amplitude of such processes is reduced by running the convolution from $\tau_1$ instead of from the start of the pulse. Naturally, asking the two photons to be emitted at the same time $\tau_1=\tau_2$ gives a vanishing amplitude; at the other extreme, the effect is not expected to be noticeable for $\tau_2-\tau_1\gg\gamma^{-1}$.

\begin{figure*}
\begin{minipage}{0.9\textwidth}
\subfloat{
\includegraphics[width=0.31\textwidth]{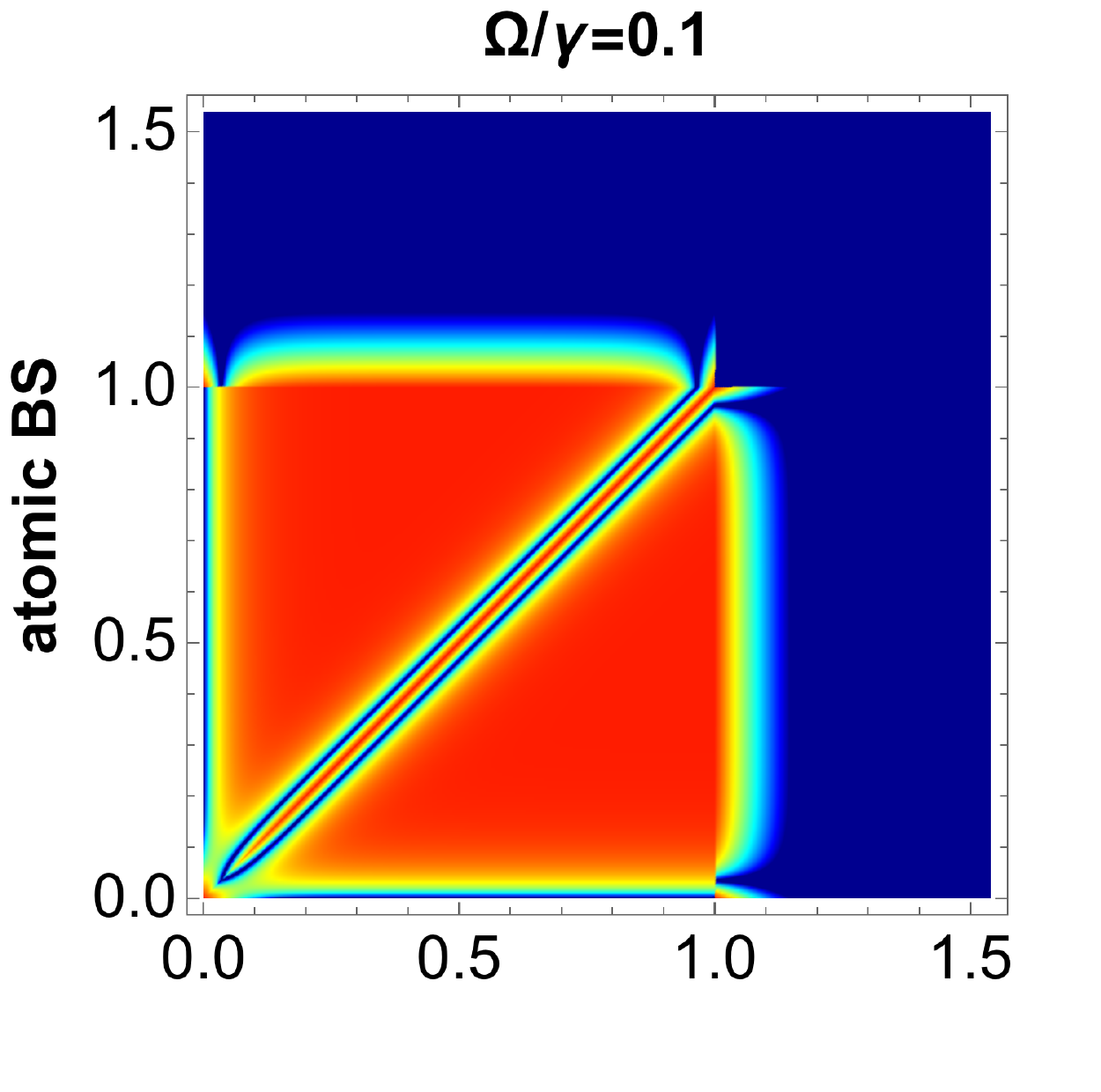}
}
\subfloat{
\includegraphics[width=0.31\textwidth]{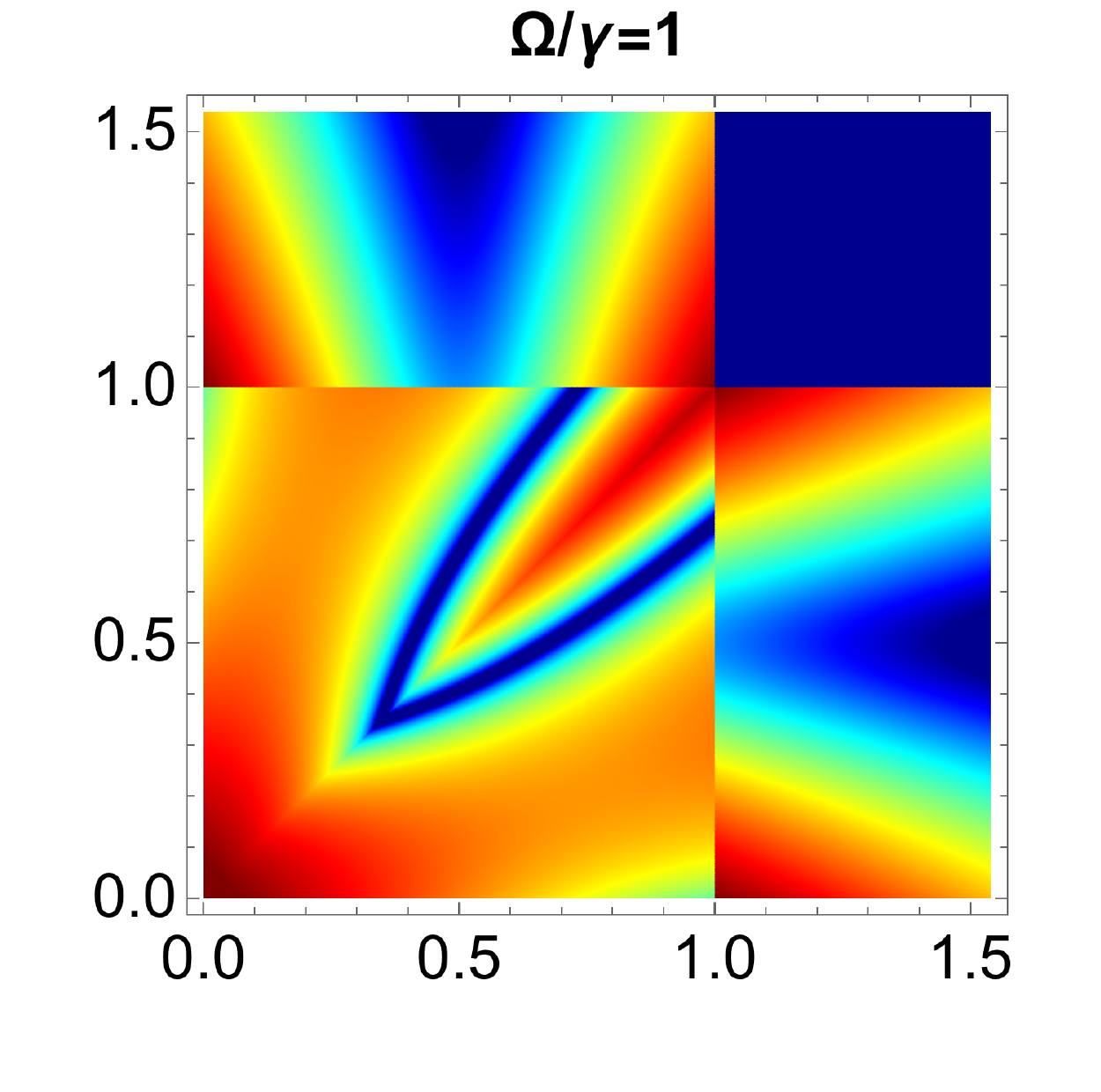}
}
\subfloat{
\includegraphics[width=0.31\textwidth]{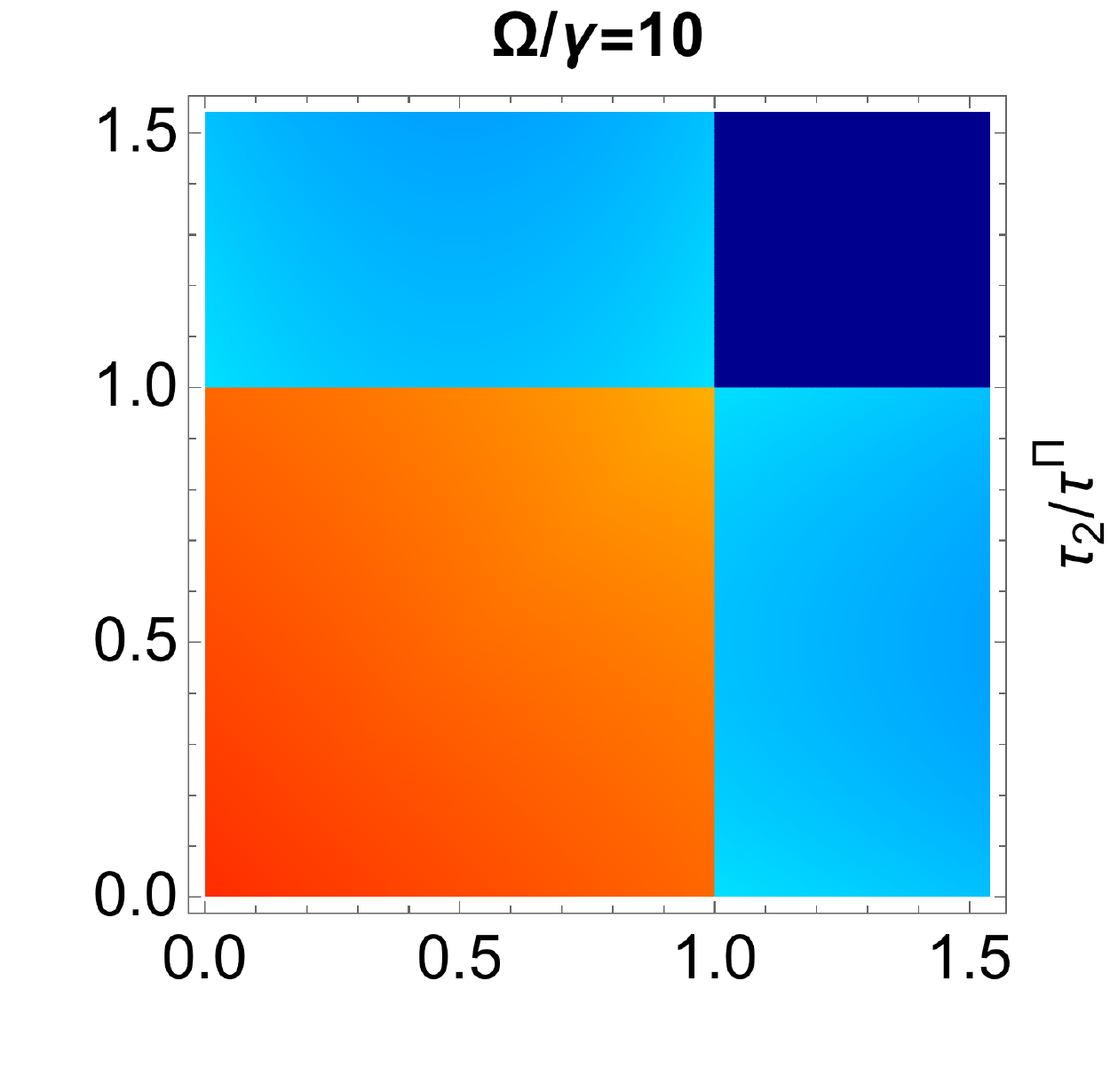}
}\\\vspace{-0.5cm}
\subfloat{
\includegraphics[width=0.31\textwidth]{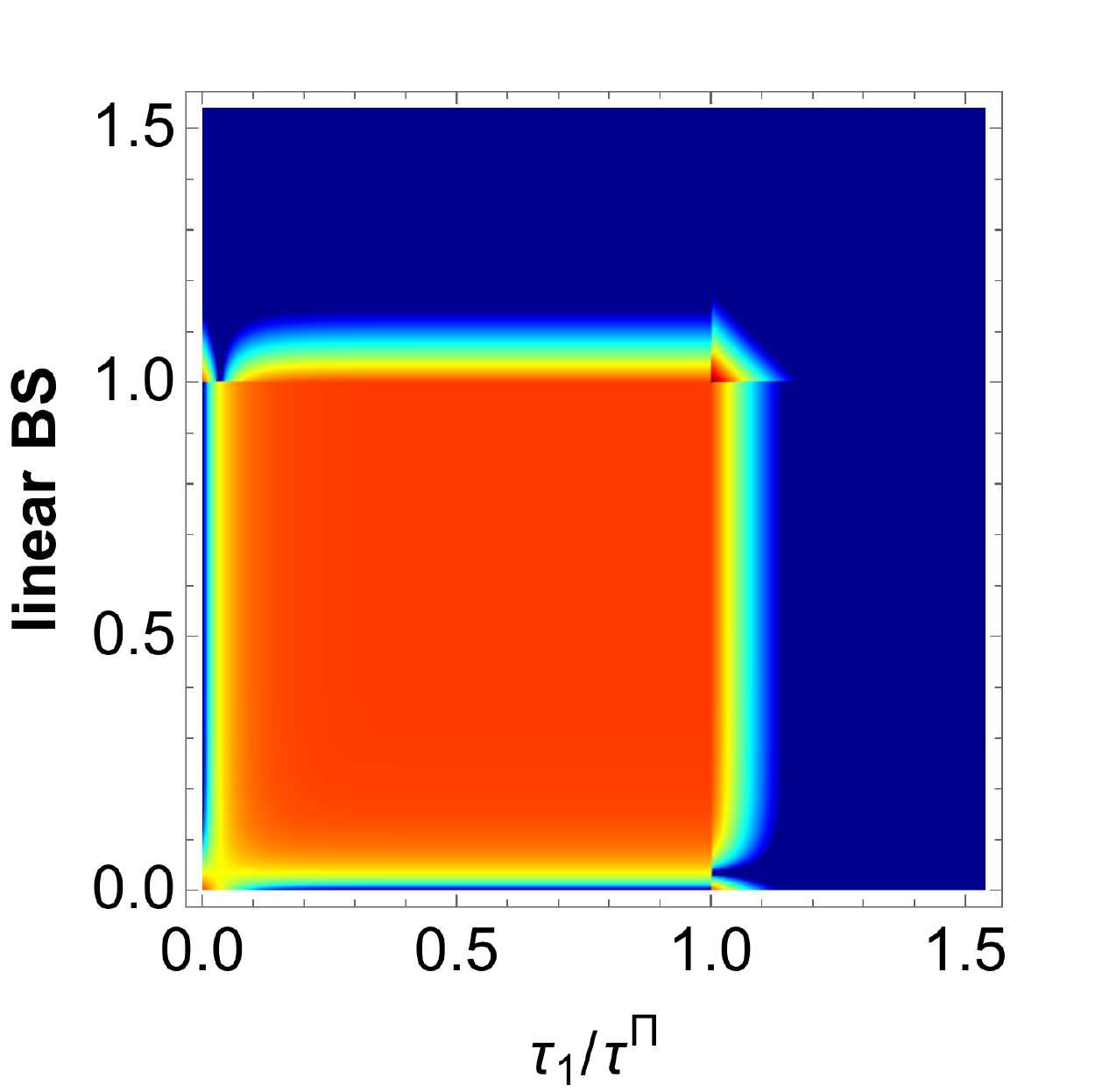}
}
\subfloat{
\includegraphics[width=0.31\textwidth]{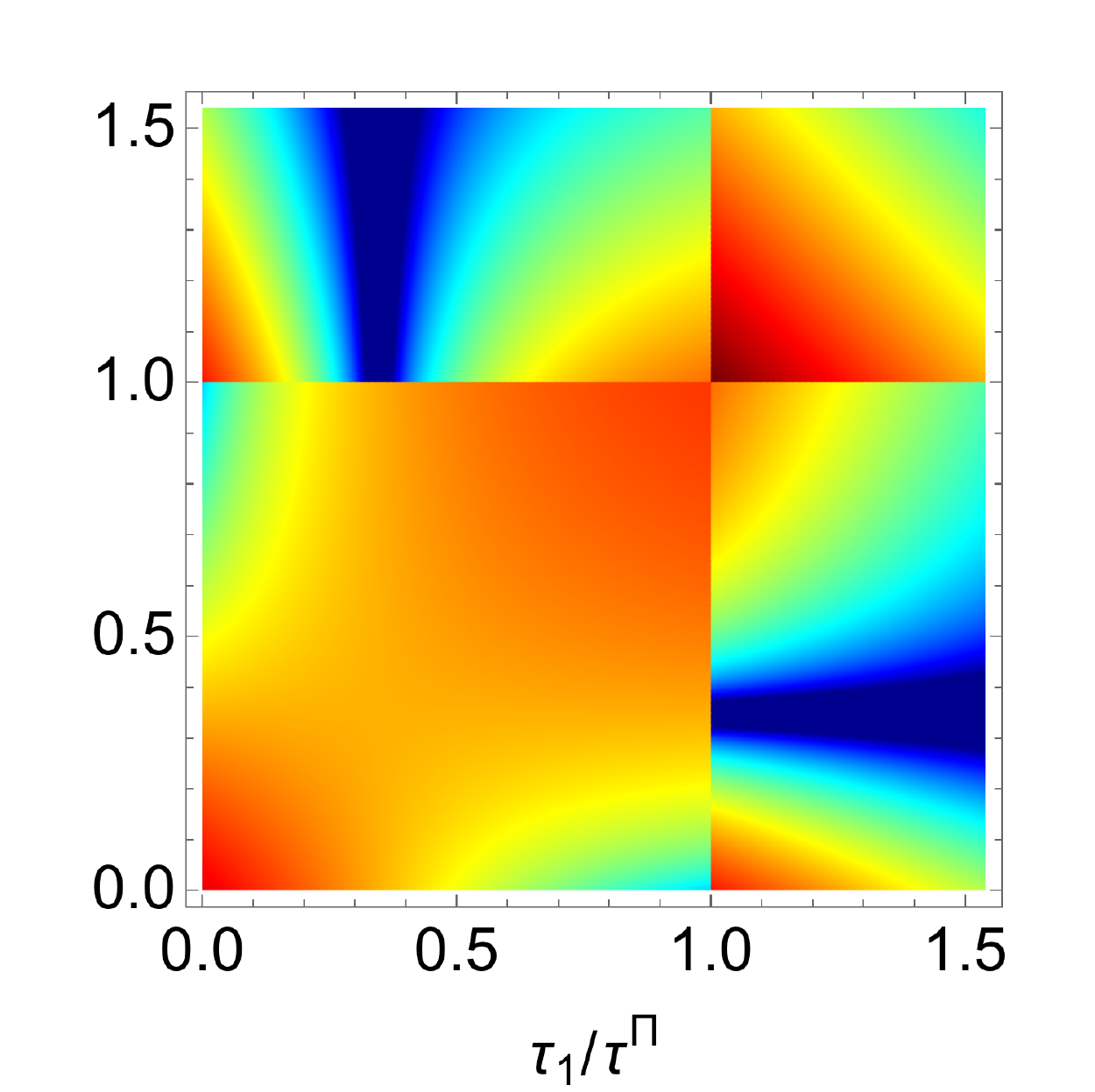}
}
\subfloat{
\includegraphics[width=0.31\textwidth]{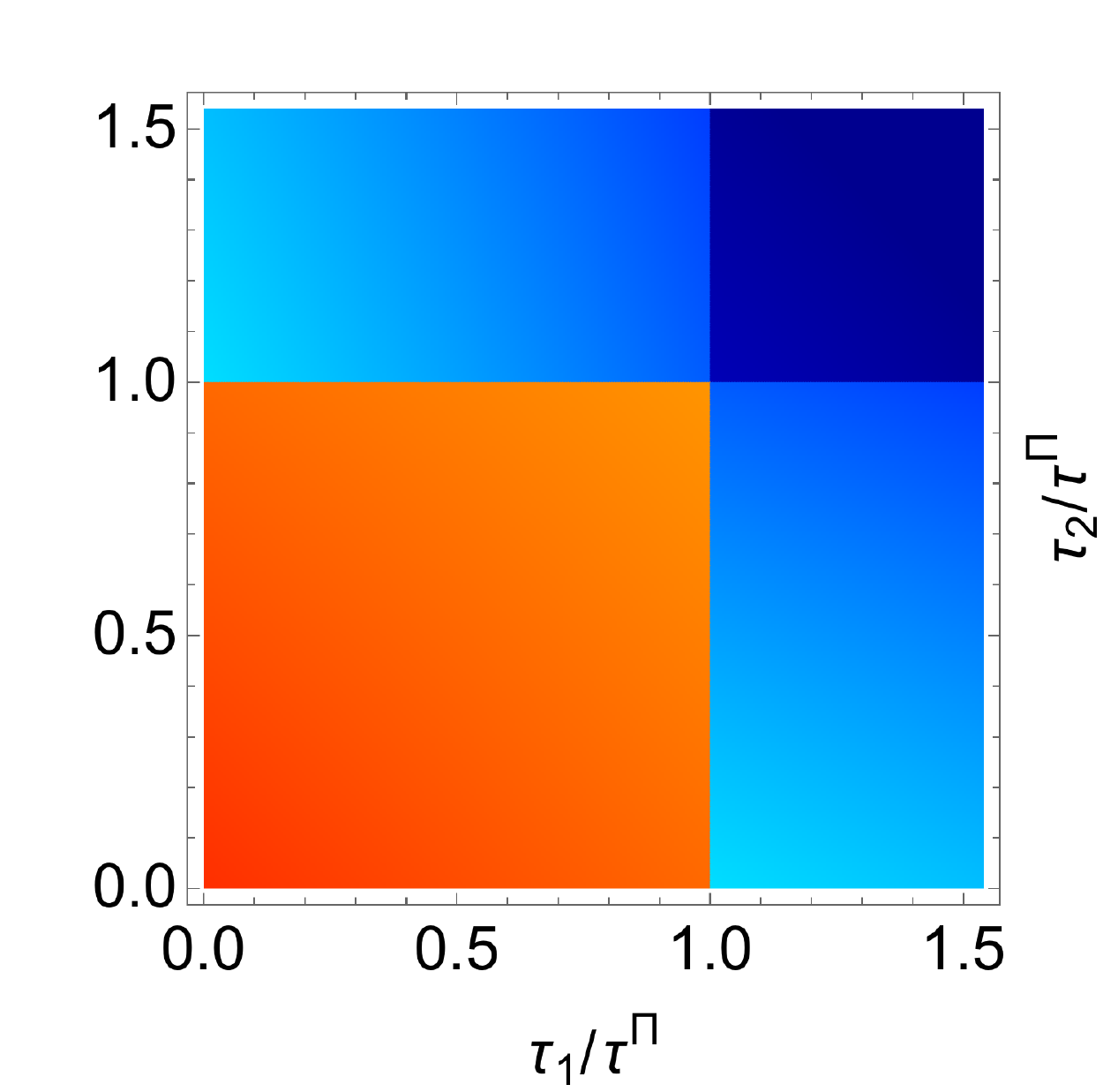}
}
\end{minipage}\begin{minipage}{0.1\textwidth}
\subfloat{
\quad\includegraphics[width=0.8\textwidth]{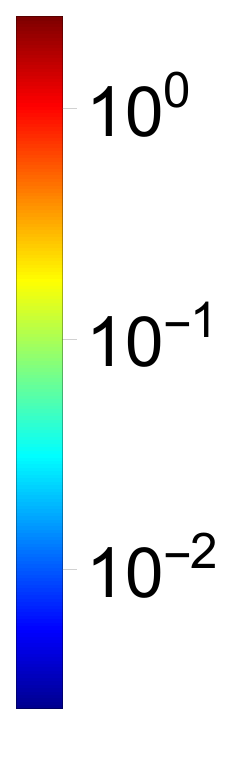}
}
\end{minipage}
\caption{\label{fig:timeDensity}(color online).\quad The time distribution of the outgoing photons post-selected on coincidence events $\lim_{t\to\infty}\mathcal{S}^{\sqcap}_{\tau_1,\tau_2}(t)/\mathcal{C}$ for resonant square pulses of various bandwidth $\Omega$. The first row corresponds to the photons being scattered by the atomic BS. Note that since the atom can only spontaneously emit a single photon, the coincidences vanish for $\tau_1,\tau_2>\tau^{\sqcap}$. The second row, meant for comparison, is the fictitious situation where the scatterer would be a linear BS with the same reflection $r_\omega$ and transmission $t_\omega$ as those of the atomic BS. The presence of valleys of zero coincidence in the region $\tau_1,\tau_2\leq\tau^{\sqcap}$ are the clearest signature of the nonlinearity induced by the atom.}
\end{figure*}

After the scattering event, the time distribution of outgoing photons, post-selected on having observed a coincidence event, is $\lim_{t\rightarrow\infty}\mathcal{S}_{\tau_1,\tau_2}(t)/\mathcal{C}$. We plot this distribution in Fig.~\ref{fig:timeDensity} for various pulse bandwidths, again both for the atomic BS (upper row) and for the hypothetical linear BS with the same frequency-dependent coefficients. The most striking evidence of non-linearity is the presence of \textit{valleys}, values of $(\tau_1,\tau_2)$ for which it is impossible to observe coincidences. As it was the case with the frequency distributions, it is not easy to comment on the intermediate regime $\Omega/\gamma=1$, but it can be explained from the two extreme cases that we analyse in detail.

Let us start with the case $\Omega/\gamma=0.1$. In the linear case, the atom is just a fully reflecting mirror, since we work at resonance, so the distribution is that of the initial square product $|\xi(\tau_1)\xi(\tau_2)|^2$, plus some features at $\tau_j \lesssim \gamma^{-1}$ and $\tau_j>\tau^{\sqcap}$ respectively due to the transient building of the atomic response and the spontaneous emission after the end of the pulse. The exact result is almost identical, but for the two valleys surrounding the diagonal $\tau_1=\tau_2$ in a window of order $\gamma^{-1}$. Interestingly, on the diagonal itself -- which is the critical situation where the nonlinearity is the strongest -- we find again the same distribution as in the linear case. These features can be explained in terms of the path decomposition given in Table~\ref{tab:pathAmp}. Let's focus on the diagonal $\tau_1=\tau_2<\tau^{\sqcap}$ first:
\begin{equation}
	\mathcal{S}_{\tau_1,\tau_1}^{\sqcap}(t>\tau_1)=\big|\xi(\tau_1)\xi(\tau_1)(-1+2e^{-\gamma\tau_1})\big|^2\ .
\end{equation}
As noted above, the vanishing nonlinear amplitude (\ref{eq:nlCorrect}) ensures that the (c) paths do not contribute. As soon as the atom is given enough time to react to one photon, \textit{i.e.} $\tau_1\gg\gamma^{-1}$, the two (b) paths fully contribute with a phase shift of $\pi$ coming from the emitted photon. The overall amplitude is therefore the same as that for the (a) path, the product of input square pulses, with a minus sign. This result might seem intruiguing given that the atom can only emit one photon at a time. Indeed, one could expect that one photon gets absorbed and reflected while the other one sees a transparent medium and goes through, which would lead to perfect bunching. Again, this is because \textit{both} (c) paths are precluded: not only the reflection of both photons, but also one of the path contributing to double transmission. Finally, the rest of the diagonal is easily understood: a dip is observed at the half-life of the atomic response function $\tau_1 = \ln(2)\gamma^{-1}$ where the (b) paths contribution is divided by two, while
only the (a) path contributes when $\tau_1\ll \gamma^{-1}$. When we move away from the diagonal, we have (assuming $\gamma^{-1}\ll\tau_1,\tau_2$)
\begin{equation}
	\mathcal{S}_{\tau_1,\tau_2}^{\sqcap}(t>\tau_1,\tau_2)=\big|\xi(\tau_1)\xi(\tau_2)(1-2e^{-\gamma|\tau_2-\tau_1|})\big|^2\ .
\end{equation}
For sufficiently separated detections $|\tau_2-\tau_1|\gg\gamma^{-1}$, the (c) paths fully contribute and compensate the (b) paths, thus leaving the initial profile given in the (a) path. In physical terms, the atom had enough time to interact with the two photons one after the other in a linear way. The two valleys correspond to $|\tau_2-\tau_1|=\ln(2)\gamma^{-1}$, where the contribution of the (c) paths has been divided by two as a consequence of the reduced time interval for the absorption of the second photon. Note that this nonlinear signature becomes more and more negligible as the bandwidth is further decreased $\Omega\ll\gamma$.

In the case $\Omega/\gamma=10$, the time distribution is barely modified by the presence of the atom. This is rather expected since, even though the input pulses are very intense in the time domain, they are mostly off-resonant and thus see the atom as an almost transparent medium. However, this was where the ``reversed HOM effect'' was the most pronounced in the frequency domain (see Fig.~\ref{fig:freqDensity}). We can now look back and show how the derivation of that effect is readily obtained from the path decomposition given in Table~\ref{tab:pathAmp}. Indeed, the condition for constructive interference in the linear regime Eq.~(\ref{eq:linConst}) corresponds to the last path in (c) being equal to the sum of all the other paths. Therefore it necessarily implies that the sum of the (a) and (b) paths vanishes. Now since the atomic BS cannot respond to both photons in this regime of short pulses $\tau^{\sqcap}\ll\gamma^{-1}$, the (c) paths do not contribute and we are left with the precise sum of (a) and (b), leading to destructive interference.

\begin{figure}
\begin{minipage}{0.4\textwidth}
\subfloat{
\includegraphics[width=1\textwidth]{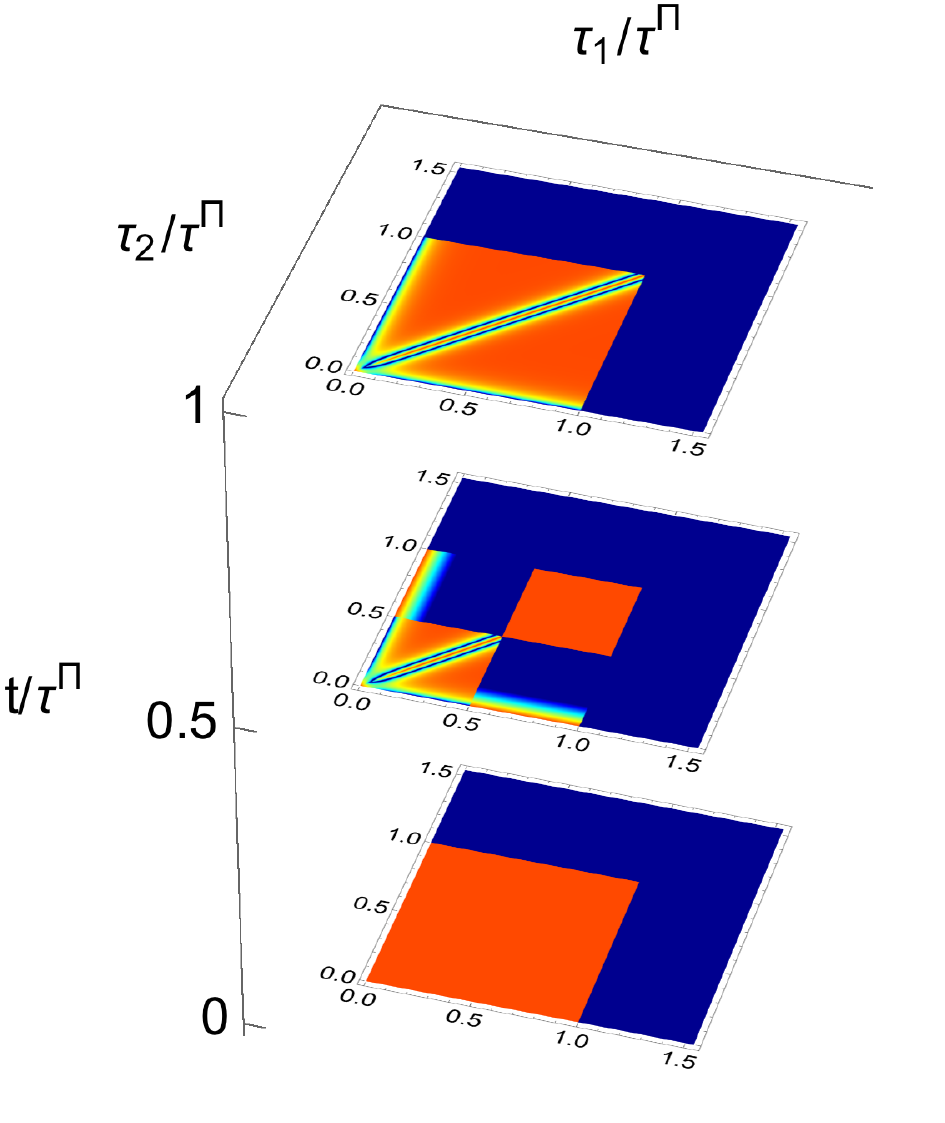}
}
\end{minipage}\begin{minipage}{0.1\textwidth}
\subfloat{
\quad\includegraphics[width=0.8\textwidth]{TimeLegend}
}
\end{minipage}
\caption{\label{fig:timeEvol}(color online).\quad Evolution of the time distribution of photons propagating in opposite directions $\mathcal{S}^{\sqcap}_{\tau_1,\tau_2}(t)$, more easily interpreted as the probability of coincidence triggered of photons at distance $\tau_j$ from the front, were the atom to be removed at time $t$. For the calculation, the input pulses are resonant with a bandwidth fixed to $\Omega/\gamma=0.1$, making the atomic BS being almost equivalent to a linear reflective mirror. If the atom is removed before the pulses arrive ($t=0$), the pulses just propagate and the coincidence shows the product of the input square profiles. If the atom is removed once the pulses have passed ($t=\tau^{\sqcap}$), we recover the corresponding result in Fig.~\ref{fig:timeDensity}, apart from the spontaneous decay of the atom that will happen for $t>\tau^{\sqcap}$. If the atom is removed half-way through the pulses ($t=\tau^{\sqcap}/2$), the distribution is divided into four distinct squares. The cases when neither photon has reached the atom yet, or when both photons have reached the atom, are analogous to the two cases above. When one of the photons has reached the atom but not the other, the feature can be explained as follows: the photon that has not reached the atom will be certainly transmitted; so, in order to observe a coincidence, the photon that has reached the atom must also have been transmitted; but this can only be a transient effect at the beginning of the scattering event, before the emission of the atom starts building up the interference which reflects the rest of the pulse. The effect thus becomes smaller and smaller as $\Omega$ is further decreased.}
\end{figure}

Finally, we note that our formalism gives us access to $\mathcal{S}_{\tau_1,\tau_2}(t)$ for finite values of the running time $t$. This represents the probability that the photons at time $\tau_j$ from their respective front are \textit{found in counter-propagating modes} at time $t$. It could also be observed as an asymptotic coincidence if one were able to suddenly remove the atom at time $t$. The changes of $\mathcal{S}_{\tau_1,\tau_2}(t)$ as a function of $t$ are presented in Fig.~\ref{fig:timeEvol} for small bandwidth $\Omega/\gamma=0.1$. The most obvious feature of the figure is the fact that photons, that had not reached the atom when it was removed ($\tau_j>t$), simply propagate. The transient coincidence observed when one photon has seen the atom but not the other, visible in the off-diagonal squares for $t=\tau^{\sqcap}/2$, is explained in the figure caption.

\section{After all, is the atomic BS integrable...}
\subsection{... as a mediator of photon-photon interaction?}
\begin{figure}
\includegraphics[width=0.46\textwidth]{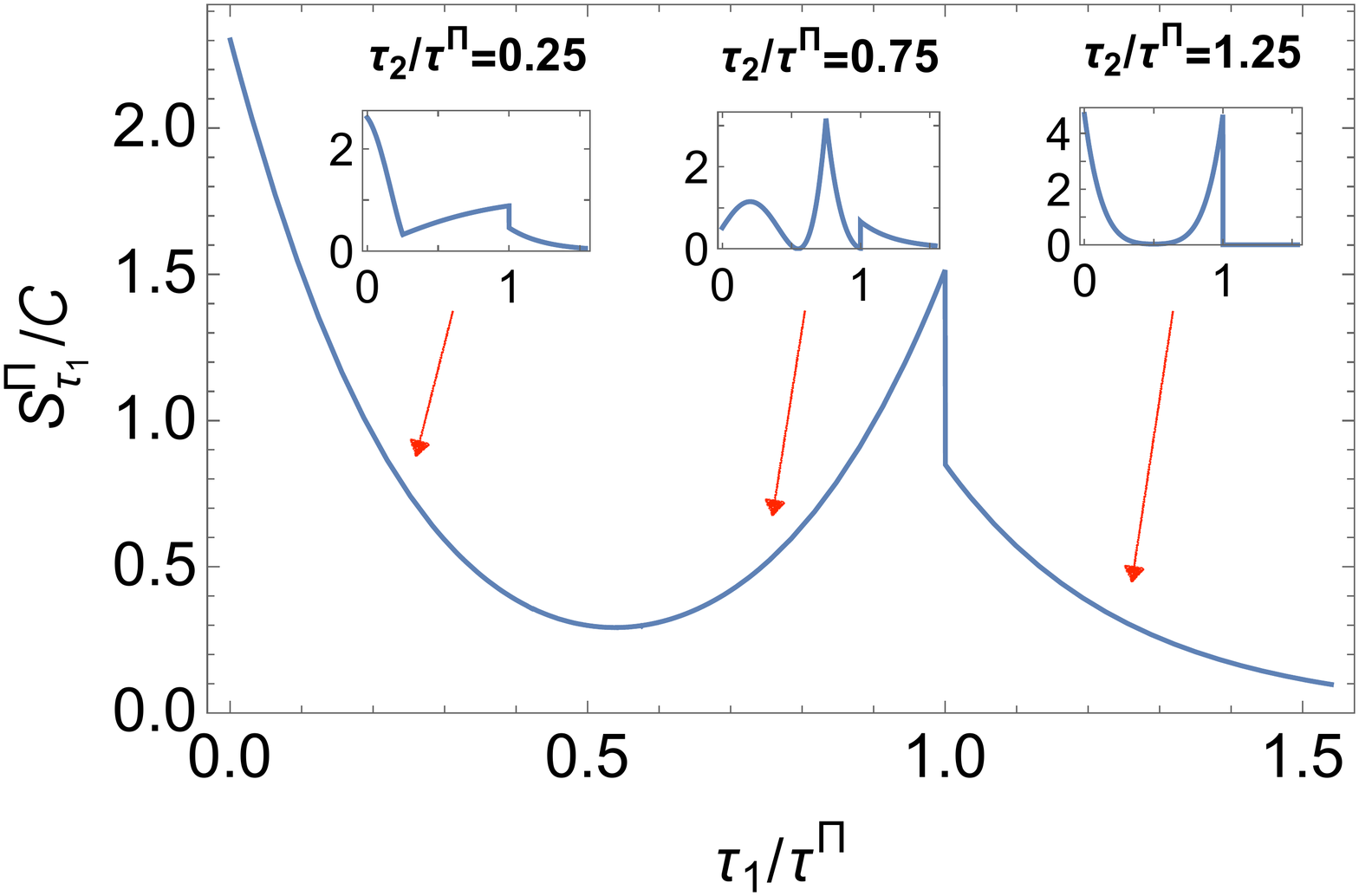}
\caption{\label{fig:shapePos} Time distribution of an outgoing photon when tracing out the other one propagating in opposite direction $\mathcal{S}^{\sqcap}_{\tau_1}/\mathcal{C}=\lim\limits_{t\to\infty}\int_{0}^{\infty}\!\mathrm{d} \tau_2\,\mathcal{S}^{\sqcap}_{\tau_1,\tau_2}(t)/\mathcal{C}$. The inset shows the time distribution of the same photon when postselecting on the detection of the other one at $\tau_2$. Here the input square pulses are resonant with a bandwidth fixed to $\Omega/\gamma=1$, corresponding to the nonlinear regime.}
\end{figure}

We now discuss the integrability of the atomic BS in the regime $\Omega/\gamma=1$ where it responds nonlinearly to co-incident photons. As mentioned previously, this nonlinearity induced by the atom could in principle allow the realization of novel practical devices \cite{Chang2014}. However, in order for the devices to be implemented in a more complex circuit, it is of utmost importance that the photons retain their initial shape and do not get mixed. To this end, Fig.~\ref{fig:shapePos} shows how the profile of one outgoing photon is correlated with the detection of the other photon propagating in opposite direction at a sharp time. Contrasting with recent results where the nonlinearity was found to decrease the correlations induced by the atom \cite{Anders2015}, we find that the outgoing photons shape is still severely modified in a non-trivial manner in this regime. This hinders the integrability of the device as a mediator of photon-photon interaction.

\subsection{... as a tunable BS?}

\begin{figure}
\subfloat{
\includegraphics[width=0.22\textwidth]{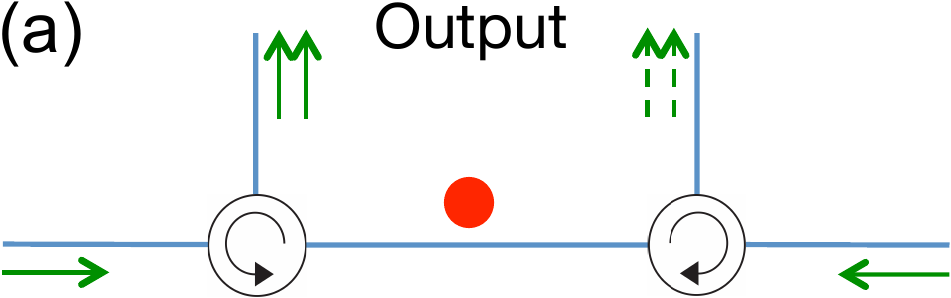}\label{fig:HOMon}
}\quad
\subfloat{
\includegraphics[width=0.22\textwidth]{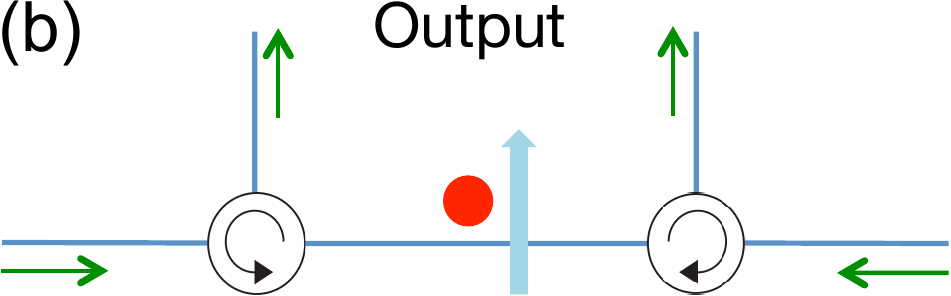}\label{fig:HOMoff}
}\\
\subfloat{
\includegraphics[width=0.40\textwidth]{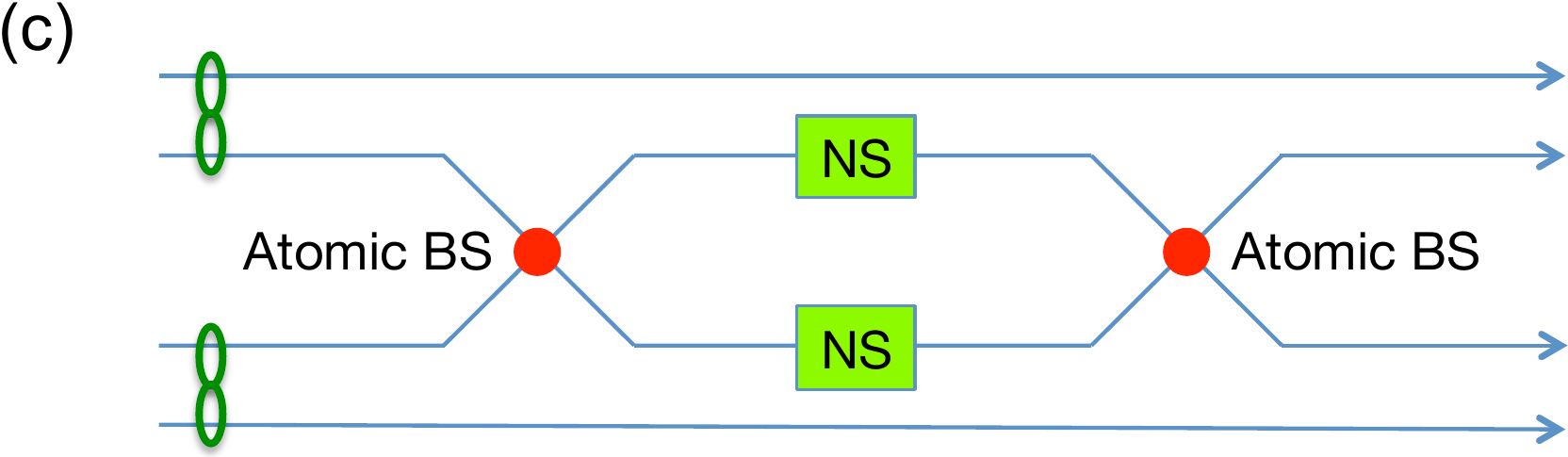}\label{fig:KLMgate}
}
\caption{\label{fig:tunableBS}(color online).\quad The atomic BS becomes four-port when the waveguide is connected to two circulators. (a) The Hong-Ou-Mandel effect is turned ON when $\Delta\approx \gamma$. (b) An external control field can turn OFF this effect by changing the detuning to $\Delta \approx 0$ or $\Delta \gg \gamma$. (c) A controlled-phase gate in the Knill-Laflamme-Milburn scheme can be switched ON or OFF with the use of the tunable atomic BS. The nonlinear sign (NS) gate implements the transformation $\alpha |0\rangle +\beta |1\rangle + \eta |2\rangle \rightarrow \alpha |0\rangle +\beta |1\rangle-\eta |2\rangle$, where a $\pi$ phase shift is applied only when the two photons reach the gate together.}
\end{figure}

For practical purposes, it is usually necessary to have a four-port BS where the input and output are in different ports, as illustrated in Fig.\,\ref{fig:glassBS}. For the waveguide-atom system this can be done by connecting the waveguide to two circulators~\cite{Sugimoto1999,Wang2005,Pintus2013,Davoyan2013,Ghosh2013} as in Fig.\,\ref{fig:tunableBS}. The outgoing photons are then rerouted to two different ports. In the linear regime,
such an atomic implementation of a BS has advantage over conventional BS because it is easily tunable, as follows from previous results~\cite{SFopt2005,Chang2007}. Indeed, we see in Eq.\,\eqref{eq:Cmono} that the two photons are totally reflected at resonance $\Delta=0$, and totally transmitted when far detuned $\Delta\gg \gamma$; in both cases there is no Hong-Ou-Mandel effect. By controlling the energy spacing of the two-level system and hence the detuning, it is possible to turn the Hong-Ou-Mandel effect ON or OFF. This can be done by adjusting the gate voltage and biased flux in the case of superconducting qubit~\cite{Wallraff2004}, or an external field for the case of atom and quantum dot~\cite{Patel2010}. Such a high-speed tunable BS is useful for feed-forward operations on photonic qubits, as discussed in Ref.~\cite{Ma2011}. In contrast with this proposal based on Mach-Zehnder interferometry, the BS under study is realized by a single atom-like two-level system, thus making it more suitable for implementation on a chip of artificial atoms. Moreover, the spectral shape of the photons is preserved given that we operate in the linear regime where the photons do not interact and are effectively monochromatic for the atomic BS.

To see an example where the tunable atomic BS can be useful, we look at the two-qubit controlled-phase gate proposed in the Knill-Laflamme-Milburn scheme for optical quantum computing~\cite{Knill2001}. The module illustrated in Fig.\,\ref{fig:KLMgate} implements such a gate on two dual-rays photonic qubits, where the 50-50 BS in the original proposal have been replaced by tunable atomic BS. When the atomic BS is in the mode ON the controlled phase-gate is implemented, but when it is in mode OFF there is no Hong-Ou-Mandel effect and one can verify that the output is the same as the input, that is, the module implements the identity operator. Such an ability to turn ON and OFF the controlled-phase gate allows the realization of a \textit{configurable} integrated optical chip that is capable of running different computational tasks, each of the tasks usually requiring a different number of controlled-phase gates to be applied. Quantum dots and superconducting qubits are natural choices for the two-level system in an implementation on an integrated chip.

\section{Conclusion}
We have presented a time-dependent study of the scattering of two photons on a quantum emitter. We have shown that in the case of quasi-monochromatic photons, a linear regime naturally arises where the atom behaves as a conventional BS, leading to a Hong-Ou-Mandel effect for the right parameters. We have also discussed a potential application of such a tunable BS, enabling to switch ON and OFF a two-qubit gate in an integrated optical chip. In addition, the nonlinearity induced by the atom has been investigated by monitoring the atomic excitation during the scattering event and a clear signature is predicted in the coincidence counts. We also explained in great detail the correlations induced by the atomic BS, studying them from a new perspective. These correlations are found to severely affect the incoming photons shape in the nonlinear regime, hindering the integrability of the device as a mediator of photon-photon interaction. The investigation of BS~\cite{Ionicioiu2011}, mirrors~\cite{Roulet2014,PO2015} and interferometers~\cite{Fratini2014PRL,Jibo2015} operating in the quantum regime opens the way to new exciting experiments, such as the quantum delayed-choice experiment~\cite{Peruzzo2012,Kaiser2012}, where controlling devices usually considered classical now behave according to quantum mechanics.

\begin{acknowledgements}
We thank Anders Nysteen for sharing with us the data of his numerical simulations and for insightful discussions. We also thank Lee Su-Yong for stimulating discussions. We are also grateful for the remarks of an anonymous referee, which have shaped significantly the final version of this manuscript. This research is supported by the National Research Foundation (partly through its Competitive Research Programme, Award No. NRF-CRP12-2013-03) and the Ministry of Education, Singapore.
\end{acknowledgements}

\appendix
\section{Derivation of the time-dependent averages}\label{sec:appA}
We give here a detailed derivation of the average coincidence \eqref{eq:Coindef} and probability of excitation of the atom \eqref{eq:Exdef} as given in the main text.

To describe the evolution of the system during the scattering event, it is convenient to work in the interaction picture with respect to the free Hamiltonian $\hat{H}_0=\hbar\omega_A |e\rangle \langle e|+\int_{0}^\infty\!d \omega\, \hbar\omega \left(\hat{a}_\omega^\dagger \hat{a}_\omega+\hat{b}_\omega^\dagger \hat{b}_\omega\right)$. The total Hamiltonian then reads
\begin{equation}\label{eq:HamiltonianInterac}
	\hat{H}_{int}=-i\hbar\int_{0}^\infty\!d \omega\, g_\omega \left[\hat{\sigma}_+ \left(\hat{a}_\omega +\hat{b}_\omega \right)e^{-i(\omega-\omega_A)t}-\mathrm{H.c.} \right] ,
\end{equation}
where $\hat{\sigma}_+=|e\rangle\langle g|$ is the atomic raising ladder operator. In the following we will use the Weisskopf-Wigner approximation \cite[p. 207]{QOptics1997}, where the coupling constant is evaluated at the transition frequency $g_{\omega}=g_{\omega_A}.$

The average coincidence \eqref{eq:Coindef} after the scattering event is $\mathcal{C}=1-P_{aa}(t \rightarrow \infty)-P_{bb}(t \rightarrow \infty)$, where $P_{jj}$ is the probability of having two photons in mode $j$. We show below how to obtain $P_{aa}$; the probability $P_{bb}$ can be computed in a similar manner. We have $P_{aa}=\bra{\psi_{in}}\hat{C}_{aa}\ket{\psi_{in}}$, where
\begin{align}\label{eq:coincidenceDeriv}
\hat{C}_{aa}=\frac{1}{2} \int_{0}^\infty\!\! d  \omega \!\int_{0}^\infty \!\! d \omega' \, \hat{a}^\dagger_{\omega} \hat{a}^\dagger_{\omega'}\hat{a}_{\omega'} \hat{a}_{\omega}=\frac{1}{2}(\hat{N}_a^2-\hat{N}_a) ,
\end{align} 
with $\hat{N}_a=\int_{0}^\infty\! d \omega\, \hat{a}_\omega^\dagger \hat{a}_\omega$ the photon-number operator in mode $a$. Note that we omit the time dependence of the field and atom operators for clarity. The Heisenberg equation of motion yields the following closed set of first-order differential equations  for the operators $\hat{C}_{aa}, \hat{N}_a, \hat{\sigma}_+,\hat{\sigma}_z, \hat{N}_a \hat{\sigma}_+,\hat{N}_a\hat{\sigma}_z$ \cite{Domokos2002,Wang2012}
\begin{widetext}
\begin{align}\label{eq:Heisenberg}
\dot{\hat{C}}_{aa}&= \frac{\gamma}{2} (\hat{N}_a+\hat{N}_a\hat{\sigma}_z)+\sqrt{\gamma}(\hat{N}_a \hat{\sigma}_+ \hat{a}_0+\mathrm{H.c.}), \\ \nonumber
\dot{\hat{N}}_a&= \frac{\gamma}{2} (\hat{\mathds{1}}+\hat{\sigma}_z)+\sqrt{\gamma}\,(\hat{\sigma}_+\hat{a}_0+\mathrm{H.c.}) ,\\ \nonumber
\dot{\hat{\sigma}}_+&=-\gamma \hat{\sigma}_+ + \sqrt{\gamma}\,(\hat{a}_0^\dagger+\hat{b}_0^\dagger)\hat{\sigma}_z, \\ \nonumber
\dot{\hat{\sigma}}_z&=-2\gamma (\hat{\mathds{1}}+\hat{\sigma}_z)- 2\sqrt{\gamma}\,[\hat{\sigma}_+(\hat{a}_0+\hat{b}_0)+\text{H.c.}], \\ \nonumber
\frac{d}{dt} \left(\hat{N}_a \hat{\sigma}_+\right)&= -\gamma \hat{N}_a\hat{\sigma}_+ + \sqrt{\gamma} \left[\hat{a}_0^\dagger(\hat{\mathds{1}}+\hat{\sigma}_z)/2 +(\hat{a}_0^\dagger+\hat{b}_0^\dagger)\hat{N}_a\hat{\sigma}_z\right] ,\\ \nonumber
\frac{d}{dt} \left(\hat{N}_a \hat{\sigma}_z\right)&= -2\gamma \left[\hat{N}_a\hat{\sigma}_z+\hat{N}_a+(\hat{\mathds{1}}+\hat{\sigma}_z)/4\right] - \sqrt{\gamma} \left[2\hat{N}_a\hat{\sigma}_+ (\hat{a}_0+\hat{b}_0)+\hat{\sigma}_+ \hat{a}_0+\mathrm{H.c.}\right] , \nonumber
\end{align}
\end{widetext}
where $\gamma=2\pi g_{\omega_A}^2$ and   $\hat{a}_0=\frac{1}{\sqrt{2\pi}}\int_{0}^\infty\!d \omega\, \hat{a}_\omega(t_0)e^{-i(\omega-\omega_A)t}$. 

To find the expectation value $\bra{\psi_{in}}\hat{C}_{aa}\ket{\psi_{in}}$ we need to know the action of the free-pulse operator $\hat{a}_0$ on the state of the system $|\psi_{in}\rangle$ at the initial time $t_0$. For this, we first express the latter in terms of creation operators as
\begin{eqnarray}\label{eqApp:in}
	|\psi_{in}\rangle&=&|1_a,1_b,g\rangle\\
	&=&\int_{0}^\infty\! d \omega\, f_a(\omega) \hat{a}^\dagger_\omega \int_{0}^\infty\! d \omega^\prime\, f_b(\omega^\prime) \hat{b}^\dagger_{\omega^\prime}|\varnothing\rangle ,\nonumber
\end{eqnarray}
where $|\varnothing\rangle\equiv|0_a,0_b,g\rangle$ corresponds to the forward and backward propagating modes being in vacuum state while the atom is in the ground state and $f_a(\omega)$ ($f_b(\omega^\prime)$) is the shape of the photon pulse incoming in mode $a$ ($b$). Specifically, we have $f_a(\omega)=f_b(\omega)$ when the two photons are indistinguishable. We then obtain
\begin{equation}
	\hat{a}_0|\psi_{in}\rangle=e^{-i\Delta t}\xi_a(t)|0_a,1_b,g\rangle,
\end{equation}
where $\xi_a(t)\equiv 1/\sqrt{2\pi}\int_{0}^\infty\!\mathrm{d} \omega\, f_a(\omega) e^{-i(\omega-\omega_0)t}$ with $\omega_0$ the central frequency of the pulse, and $\Delta = \omega_0-\omega_A$.
The free-pulse operator $\hat{a}_0$ thus decreases the number of photons when applied on $|\psi_{in}\rangle$. 

Using the first equation in Eq.\,(\ref{eq:Heisenberg}), we can now derive a differential equation for $\bra{\psi_{in}}\hat{C}_{aa}\ket{\psi_{in}}$ 
\begin{eqnarray}
\frac{d}{dt}&&\bra{\psi_{in}} \hat{C}_{aa}\ket{\psi_{in}}=\frac{\gamma}{2} \left(\bra{\psi_{in}}\hat{N}_a\ket{\psi_{in}}+\bra{\psi_{in}}\hat{N_a}\hat{\sigma}_z\ket{\psi_{in}}\right) \nonumber \\ 
&&+\sqrt{\gamma}\left(\bra{\psi_{in}}\hat{N}_a \hat{\sigma}_+\ket{0_a,1_b,g}e^{-i\Delta t}\xi_a(t)+\text{c.c.}\right).
\end{eqnarray}
We then continue to use Eq.\,(\ref{eq:Heisenberg}) to find the differential equations for the expectation values that appear on the RHS of the above equation.

Iterating this procedure gives a system of 19 first-order differential equations, in which the only time dependence on the RHS is given by the input pulses $\xi_a(t)$ and $\xi_b(t)$. This happens because the operators $\hat{a}_0,\hat{b}_0$ are placed on the right, and $\hat{a}_0^\dagger,\hat{b}_0^\dagger$ on the left of every terms. Therefore, the number of photons is always decreased and one eventually ends up with averages in the vacuum state $\bra{\varnothing}\hat{O}\ket{\varnothing}$ with $\hat{O}$ one of the operators whose derivative is given on the LHS of Eq.\,(\ref{eq:Heisenberg}). These quantities are easily known since the system does not evolve if the state is in the vacuum $\ket{\varnothing}$. We observe that the final system of first-order differential equations can be solved one by one, which greatly simplifies the computation for which we used the software Mathematica.

In the process described above, one is led to solve the differential equation
\begin{eqnarray}
	\frac{d}{dt}&&\bra{\psi_{in}} \hat{\sigma}_{z}\ket{\psi_{in}}=-2\gamma \left(1+\bra{\psi_{in}}\hat{\sigma}_{z}\ket{\psi_{in}}\right) \\ 
	&&-2\sqrt{\gamma}\left(\bra{\psi_{in}}\hat{\sigma}_+\ket{0_a,1_b,g}e^{-i\Delta t}\xi_a(t)+\text{c.c.}\right)\nonumber \\
	&&-2\sqrt{\gamma}\left(\bra{\psi_{in}}\hat{\sigma}_+\ket{1_a,0_b,g}e^{-i\Delta t}\xi_b(t)+\text{c.c.}\right) ,\nonumber
\end{eqnarray}
which gives the probability of excitation \eqref{eq:Exdef}.

We now have all the necessary information to study the problem of two-photon bunching on the atom for square pulses of the form
\begin{equation}
	\xi_a(t)=\xi_b(t+T)=\left\{
    \begin{array}{cl}
        \sqrt{\frac{\Omega}{2}} \qquad& \text{for } t_0\leq t\leq \frac{2}{\Omega} \\
        0\quad & \text{otherwise}
    \end{array}
\right. ,
\end{equation}
where $T$ represents the delay between the two pulses incoming on the BS. The monochromatic regime then corresponds to the limit $\Omega\ll\gamma$ and is independent of the exact pulse shape.

Specifically, the probability of excitation during the time interval $t_0\leq t\leq \frac{2}{\Omega}$ is found to be
\begin{widetext}
	\begin{eqnarray}
	\frac{\bra{\psi_{in}} \hat{\sigma}_{z}\ket{\psi_{in}}+1}{2}&=& \frac{\sigma  e^{-2 t^\prime}}{\delta  \left(\delta ^2+1\right)^3} \Big[\delta  \left(2 \sigma  \left((2 t^\prime-3) \delta ^2+2
		   t^\prime+5\right)+\left(\delta ^2+1\right)^2\right) +e^{2  t^\prime } \left(\delta ^3+\delta \right) \left(-2 \sigma
		   +\delta ^2+1\right) \\
		   &&-2 e^{t^\prime} \delta 
		   \left(\left(\delta ^2+1\right)^2-2 \sigma  \left(( t^\prime +2) \delta ^2+ t^\prime -2\right)\right)
		   \cos ( t^\prime  \delta )+4 \sigma  e^{ t^\prime } \left( t^\prime  \delta ^4+( t^\prime -3)
		   \delta ^2+1\right) \sin ( t^\prime  \delta )\Big] ,\nonumber
	\end{eqnarray}
\end{widetext}
where $t^\prime=\gamma (t-t_0)$, $\delta=\Delta/\gamma$ and $\sigma=\Omega/\gamma$ are respectively the normalized time, detuning and bandwidth. Eq.\,(\ref{eq:Exlin}) is then readily obtained by considering the monochromatic regime $\sigma\ll 1$ while Eq.\,(\ref{eq:Exnonlin}) corresponds to the resonant case $\delta=0$.

The solution for the average coincidence $\mathcal{C}$ is rather lengthy, but is greatly simplified in the monochromatic limit (Eq.\,(\ref{eq:Cmono})) or at resonance (Eq.\,(\ref{eq:Cnonlin})).

We also plot in Fig.\,\ref{fig:Cnonlin} the average coincidence $\mathcal{C}$ for Gaussian pulses of the form $\xi_a(t)=\xi_b(t)=\left(\frac{4\Omega^2}{\pi}\right)^{1/4} e^{-2\Omega^2t^2}$ as well as exponentially rising pulses
\begin{equation}
	\xi_a(t)=\xi_b(t)=\left\{
    \begin{array}{cl}
        \sqrt{\Omega}\,e^{\frac{\Omega}{2}t} \qquad& \text{for } t< 0 \\
        0\qquad& \text{for } t> 0
    \end{array}
\right. .
\end{equation}

\section{Derivation of the time-dependent distributions $\mathcal{S}_{\omega_1,\omega_2}(t)$ and $\mathcal{S}_{\tau_1,\tau_2}(t)$}\label{sec:appB}

The goal of this section is to compute the joint spectral decomposition given in Eq.~\eqref{eq:nonlinearSpec} and the two-time correlation in Eq.~\eqref{eq:nonlinearTime}. To this end, we use the same method as in Appendix \ref{sec:appA} and derive a closed set of equations for the amplitude $\langle 0_a,0_b,g|\hat{a}_{\omega_1}(t)\hat{b}_{\omega_2}(t)|\psi_{in}\rangle$. Omitting the time dependence of the field and atom variables, we obtain at resonance $\Delta=0$
\begin{eqnarray}\label{eq:csFreq}
	\dot{c}_{s}(\omega_1,\omega_2)&=&\sqrt{\frac{\gamma}{2\pi}}e^{i(\omega_2-\omega_A)t}c_{a\sigma}(\omega_1)+1\leftrightarrow 2\nonumber \\
	\dot{c}_{a\sigma}(\omega)&=&-\gamma c_{a\sigma}(\omega)-2\sqrt{\gamma}\xi(t)c_{a}(\omega) \\
	\dot{c}_a(\omega)&=&\sqrt{\frac{\gamma}{2\pi}}e^{i(\omega-\omega_A)t} c_A \nonumber\\
	\dot{c}_A&=&-\gamma c_A-\sqrt{\gamma}\xi(t)\nonumber ,
\end{eqnarray}
where we have used $\dot{\hat{a}}_\omega=\sqrt{\gamma/2\pi} e^{i(\omega-\omega_A)t} \hat{\sigma}_-$, Eqs.~\eqref{eq:Heisenberg} and defined the amplitudes as follows
\begin{eqnarray}
	c_{s}(\omega_1,\omega_2)&\equiv&\langle 0_a,0_b,g|\hat{a}_{\omega_1}\hat{b}_{\omega_2}|\psi_{in}\rangle\\
	c_{a\sigma}(\omega)&\equiv&\langle 0_a,0_b,g|\hat{a}_{\omega}\hat{\sigma}_-|\psi_{in}\rangle \nonumber\\
	c_{a}(\omega)&\equiv&\langle 0_a,0_b,g|\hat{a}_{\omega}(|1_a,0_b,g\rangle+|0_a,1_b,g\rangle)/2 \nonumber\\
	c_{A}&\equiv&\langle 0_a,0_b,g|\hat{\sigma}_-(|1_a,0_b,g\rangle+|0_a,1_b,g\rangle)/2\nonumber .
\end{eqnarray}
One can easily check by formally integrating the last equation of \eqref{eq:csFreq} that $c_{A}$ as defined here is precisely the linear amplitude introduced in Eq.~\eqref{eq:linAmp}.

The last step in order to further simplify the set of Eqs.~\eqref{eq:csFreq} is to take its Fourier transform, which reads
\begin{eqnarray}\label{eq:csTime}
	\dot{c}_{s}(\tau_1,\tau_2)&=&\sqrt{\gamma}\delta(t-\tau_2)c_{a\sigma}(\tau_1)+1\leftrightarrow 2\nonumber \\
	\dot{c}_{a\sigma}(\tau)&=&-\gamma c_{a\sigma}(\tau)-2\sqrt{\gamma}\xi(t)c_{a}(\tau) \\
	\dot{c}_a(\tau)&=&\sqrt{\gamma}\delta(t-\tau) c_A \nonumber\\
	\dot{c}_A&=&-\gamma c_A-\sqrt{\gamma}\xi(t)\nonumber .
\end{eqnarray}
From here, the two-time correlation $\mathcal{S}_{\tau_1,\tau_2}=|c_{s}(\tau_1,\tau_2)|^2$ given in Eq.~\eqref{eq:nonlinearTime} follows directly from integrating Eqs.~\eqref{eq:csTime}. Taking the inverse Fourier transform of the obtained $c_{s}(\tau_1,\tau_2)$, one can also check that we find the joint spectral distribution after the scattering event \eqref{eq:nonlinearSpec}, which agrees with post-scattering theories such as Refs.~\cite{SFSchwingerPRL2007,Anders2015}.

\section{Induced non-linearity for the off-resonant case $|\Delta|=\gamma$}
\label{appoffres}

\begin{figure}
\includegraphics[width=0.46\textwidth]{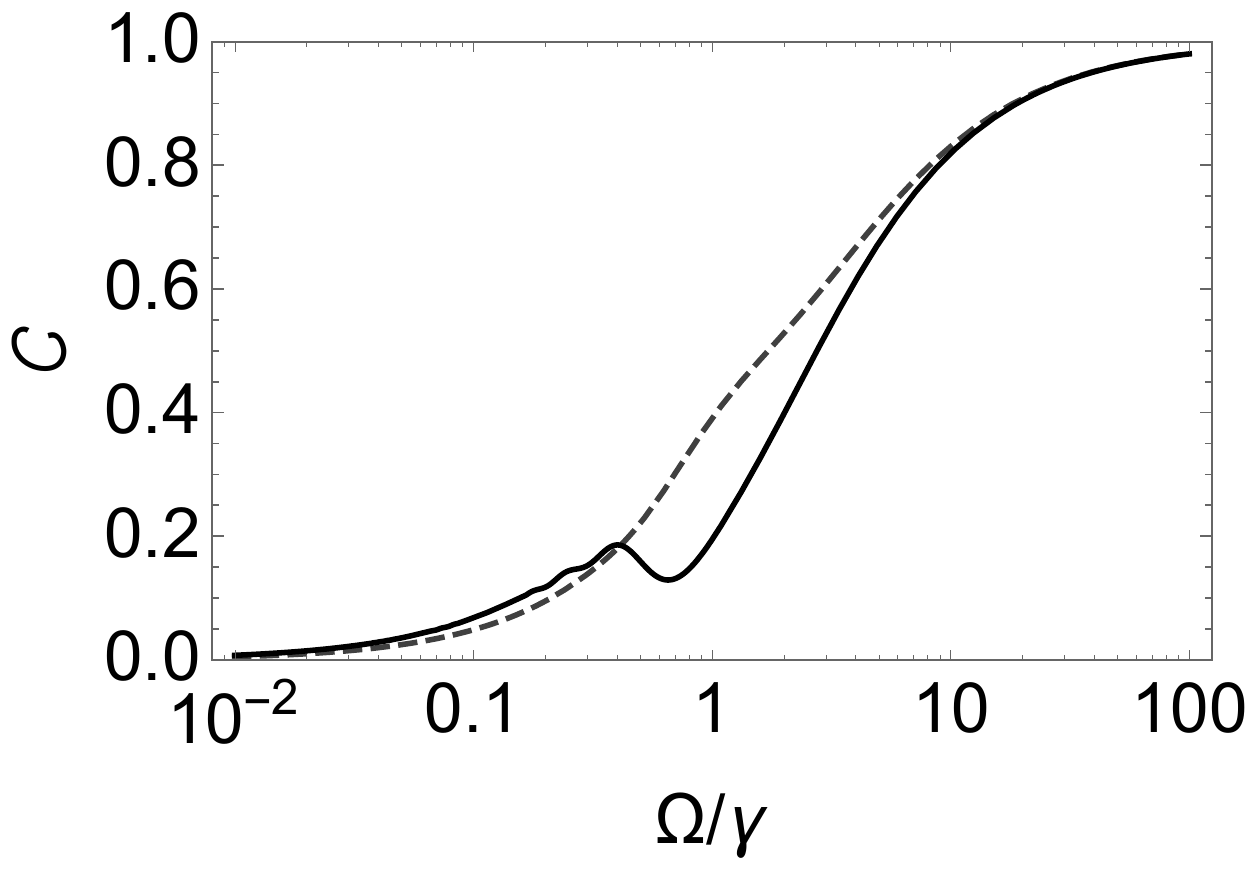}
\caption{\label{Sfig:Cnonlin} Same as Fig.\,\ref{fig:Cnonlin}, for $|\Delta|=\gamma$ such that the atomic BS acts as a 50-50 BS in the monochromatic regime.}
\end{figure}

For completeness, we briefly discuss the induced nonlinearity in the off-resonant situation $|\Delta|=\gamma$ for which the atomic BS acts as a 50-50 BS in the monochromatic regime (see Fig.\,\ref{fig:Cmono}). We omit here the analytical formula which is rather lengthy and does not provide any further understanding. Instead, Fig.\,\ref{Sfig:Cnonlin} illustrates the coincidence for square pulses of finite bandwidth $\Omega$. We observe as well a significant deviation between the atomic BS and a linear BS in the regime where $\Omega\approx\gamma$, as in the resonant case.

\bibliography{Bibliography}

\end{document}